\documentclass[onecolumn]{IEEEtran}

\ifCLASSINFOpdf
  \usepackage[pdftex]{graphicx}
 \else
  \usepackage[dvips]{graphicx}
\fi

\usepackage[cmex10]{amsmath}
\usepackage{amssymb}

\usepackage{array}

\usepackage{amssymb}
\usepackage{enumitem}
\usepackage{verbatim}
\usepackage{bbold}

\newtheorem{thm}{Theorem}

\newtheorem{cor}{Corollary}
\newtheorem{lem}{Lemma}

\newtheorem{rem}{Remark} 

\newcommand{\E}{\mathbb{E}}

\newcommand{\beq}{\begin{equation}}
\newcommand{\eeq}{\end{equation}}
\newcommand{\var}{\text{Var}}

\makeatletter

\newcommand{\Rmnum}[1]{\expandafter\@slowromancap\romannumeral #1@}
\makeatother

\begin{document}
\title{Bounds on Variance for Unimodal Distributions}

\author{Hye Won Chung,~Brian M. Sadler,~and Alfred O. Hero% \thanks{Manuscript received January 20, 2002; revise d January 30, 2002. This work was supported by the I EEE.}%
\thanks{Hye Won Chung (hwchung@kaist.ac.kr) is with the School of Electrical Engineering at KAIST in South Korea. Alfred O. Hero (hero@eecs.umich.edu) is with the EECS department at the University of Michigan. Brian M. Sadler (brian.m.sadler6.civ@mail.mil) is with the US Army Research Laboratory. This research was supported in part by ARO grants W911NF-11-1-0391 and W911NF-15-1-0479.
}}

\maketitle
%\vskip0.25in

\begin{abstract}

We show a direct relationship between the variance and the differential entropy for  subclasses of symmetric and asymmetric unimodal distributions by providing an upper bound on variance in terms of entropy power.
Combining this bound with the well-known entropy power lower bound on variance, we prove that the variance of the appropriate subclasses of unimodal distributions can be bounded below and above by the scaled entropy power. As differential entropy decreases, the variance is sandwiched between two exponentially decreasing functions in the differential entropy. 
This establishes that for the subclasses of unimodal distributions, the differential entropy can be used as a surrogate for concentration of the distribution. 

\end{abstract}


\IEEEpeerreviewmaketitle

\section{Introduction}

In this paper, we establish a direct relationship between variance and differential entropy of subclasses of symmetric and asymmetric unimodal distributions over $\mathbb{R}$.
The variance of a random variable $X$ having a distribution with density function $p(x)$ with mean $m\in \mathbb{R}$ is denoted by
\beq\label{eqn:1var}
\text{Var}(X)=\int_{-\infty}^{\infty} (x-m)^2 p(x) dx,
\eeq
and the differential entropy  of $X$ with density function $p(x)$  by
\beq\label{eqn:2ent}
h(p)=-\int_{-\infty}^{\infty} p(x)\log p(x) dx.
\eeq
Our main contribution in this paper is the observation that for appropriate subclasses of unimodal distributions, the variance of $p(x)$ can be bounded as
\beq\label{eqn:general11}
\frac{e^{2h(p)}}{2\pi e} \leq  {\text{Var}}(X) \leq \frac{c e^{2h(p)} }{2\pi e}
\eeq
for some  constant $c\geq 1$. Thus, for such  unimodal densities, as $h(p)\to -\infty$, 
the variance is guaranteed to converge to $0$.

The lower bound in~\eqref{eqn:general11} on variance  is a well-known result and holds for all probability densities. This is established by the estimation counterpart to Fano's inequality (Theorem 8.6.6 in~\cite{cover2012elements})%, and the equality is achieved  for Gaussian distributions. 
\beq\label{eqn:est_counter_Fano}
 \frac{e^{2h(p)}}{2\pi e}\leq \text{Var}(X),
\eeq
where equality is achieved for Gaussian distributions. 
%Therefore, for Gaussian distributions, there indeed exists a monotonic relationship between variance and differential entropy. 
This inequality shows that for general distributions, variance can approach 0 only if its differential entropy converges to $-\infty$.

The question is then whether there exists a generally applicable upper bound on variance in terms of differential entropy for all probability densities. The answer is negative and we can easily construct a counterexample. 
Consider the distribution 
\beq
p(x)=\left\{ 
  \begin{array}{l l}
    \epsilon, & \quad x\in \left[-t-\frac{1}{2\epsilon},-t\right]\cup \left[t,t+\frac{1}{2\epsilon}\right],\\
    0, & \quad \text{otherwise}.
  \end{array} \right.
\eeq
The differential entropy of this distribution is $h(p)=\log\frac{1}{\epsilon}$, which is independent of $t$, but the variance  is $\text{Var}(X)=t^2+\frac{t}{2\epsilon}+\frac{1}{12\epsilon^2}$, which increases without bound in $t$. Thus in general there does not exist an upper bound on  variance that is monotone in differential entropy, and, even if  the differential entropy of a distribution goes to $-\infty$, the variance of this distribution can be strictly larger than a positive constant.

However, for certain distributions, including Gaussian  and uniform, there does exist a monotonic relationship between variance and differential entropy. 
For a Gaussian distribution with mean $m$ and variance $\sigma^2$, denoted $\mathcal{N}(m,\sigma^2)$, the entropy power, defined as $e^{2h(p)}$, is proportional to the variance as
\beq\label{eqn:nice}
\sigma^2=\frac{e^{2h(p)}}{2\pi e}.
\eeq 
For a uniform distribution $ p(x)=\mathrm{unif}(m-\frac{1}{2\epsilon},m+\frac{1}{2\epsilon})$, the variance is equal to $1/(12\epsilon^2)$ and the differential entropy is $\log(1/\epsilon)$.
Thus, for uniform distributions, we have
\beq
\text{Var}\left(X\right)=\frac{e^{2h(p)}}{12}.
\eeq
Therefore, for these cases, the variance is proportional to the entropy power $e^{2h(p)}$. %decreases exponentially as $h(p)$ decreases.

Finding such a monotonic relationship between variance and differential entropy for a larger class of distributions than the Gaussian or uniform is of broad interest with many applications in signal processing, machine learning, information theory, probability, and statistics. For instance, in target localization or state estimation problems, differential entropy has been often used as a natural objective for developing sensor selection or querying strategies to minimize estimation errors~\cite{jedynak2012twenty,tsiligkaridis2014collaborative,wang2004entropy}. For example, in~\cite{wang2004entropy} an entropy-based sensor selection heuristic for target localization was proposed, which selects the most informative sensor that would yield on average the greatest or nearly the greatest reduction in the entropy of the target location distribution. 
The effectiveness of this heuristic was evaluated using simulations with Gaussian sensing models. 
However, there are few theoretical guarantees for the performance of entropy-based policies for the non-Gaussian case, where a monotonic relationship between variance and entropy is unavailable or unknown.
%However, there have been few available theoretical guarantees for the performance of such entropy-based policies, especially for sensing models other than Gaussians for which a nice monotonic relationship between variance and entropy such as~\eqref{eqn:nice} does not hold any more. 
Therefore, it is of great interest to find bounds of the form~\eqref{eqn:general11} for a more general set of densities than the Gaussian in order to provide theoretical guarantees for diverse entropy-based tasks such as waveform design or sensor selection.

In probability and statistics, there have been studies to find the most general classes of distributions that resemble Gaussian distributions, for which several functional inequalities hold. % that hold for Gaussians also hold. 
In~\cite{bakry2008simple,bobkov1999isoperimetric,madiman2016forward}, an important observation was made that several functional inequalities that hold for Gaussian, such as Poincare and logarithmic Sobolev inequalities as well as reverse entropy power inequalities, also hold for a random variable $X$ whose density function $p(x)$ is log concave, i.e., 
\beq
p(\alpha x+(1-\alpha)y)\geq p(x)^{\alpha}p(y)^{1-\alpha}
\eeq
for each $x,y\in\mathbb{R}$ and each $0\leq \alpha\leq 1$.
In particular, in~\cite{bobkov2011entropy}, it was shown that for log-concave distributions the variance of $p(x)$ can be bounded  in terms of the entropy power, $e^{2h(p)}$, as
\beq\label{eqn:des_ineq}
\frac{e^{2h(p)}}{2\pi e} \leq \text{Var}(X)\leq c_0  e^{2h(p)},
\eeq
for some positive constant $c_0<2$. By using~\cite{hensley1980slicing,webb1996central}, a tighter upper bound, which makes $c_0$ as small as 1/2, can  be shown. %it can be shown that the constant $c_0$ is as small as 1/2.
In recent work~\cite{marsiglietti2017lower}, it was also shown that the entropy and $p$-th absolute moment of a symmetric log-concave random variable are comparable.
The fact that log-concave densities resemble Gaussian has led to several applications of log-concave densities in inference and modeling~\cite{walther2009inference, bagnoli2005log,saumard2014log}.
%This result is important in the sense that even though there is no monotonic relation between variance and entropy, for any log-concave distribution variance is guaranteed to approach 0 if and only if its differential entropy converges to $-\infty$.

In this paper, we establish the complementary result that resemblance to Gaussian can be extended to subclasses of unimodal distributions that even includes some heavy-tailed distributions that are not log-concave. More precisely, we provide bounds of the form~\eqref{eqn:general11} on the variance of subclasses of unimodal densities.
%provide bounds on variance such an upper bound on variance in terms of the entropy power extends to the general class of symmetric unimodal distributions. 
%(definition of symmetric unimodal distribution and sorts including non log-concave distribution)
These subclasses include symmetric unimodal linear mixture densities of the form
\beq
p(x)=\sum_{i=1}^n \alpha_i p_i(x)\quad\text{for}\;\; \alpha_i>0,\;\;\sum_{i=1}^n \alpha_i=1,
\eeq
that is a mixture of exponentially decreasing distributions, $p_i(x)\propto e^{-\beta_i|x-m|^{\theta_i}}$ with any $\beta_i>0$, $\theta_i>0$, or uniform distributions, $p_i(x)=\mathrm{unif}\left(m-\frac{1}{2\epsilon_i},m+\frac{1}{2\epsilon_i}\right)$ with any $\epsilon_i>0$, for $i=1,\dots,n$. 
Also, we establish variance bounds for more general (not necessarily linear mixture or symmetric) unimodal densities $p(x)$ with bounded support $[b-s_l,b+s_r]$ having the unique mode at $x=b$ and satisfying Lipschitz continuity with constant $c_s>0$, i.e.,
\beq\label{eqn:Lipschitz_cl0}
|p(x+y)-p(x)| \leq c_s |y|
\eeq
for any $x, y\in[b-s_l,b+s_r]$.
There exist many unimodal distributions considered in this paper that are not log-concave, e.g., the class of generalized Gaussian densities 
\beq\label{eqn:single_unimode1}
p(x)=\frac{1}{Z(\theta, \beta)} e^{-\beta|x-m|^\theta},\;\text{ for } \beta>0, 
\eeq
with order $0<\theta<1$ where  $Z$ is a normalizing constant: $Z(\theta,\beta)=\int_{-\infty}^{\infty}  e^{-\beta|x-m|^\theta}dx$. Note that these include some heavy-tailed distributions.

Unimodal distributions and Gaussian mixtures have been widely studied and used in probability theory, statistics, signal processing and machine learning, and in particular, for estimation and testing~\cite{rao1969estimation, wegman1970maximum, groeneboom1984estimating, birge1997estimation, hartigan1985dip,eskenazis2016gaussian}.
Thus our extended entropy upper bound on variance may have broad applicability. For example, in Bayesian sequential optimal design of experiments~\cite{degroot1962uncertainty,geman1996active,tsiligkaridis2014collaborative}, successive entropy minimization is often proposed as a way to progressively concentrate the posterior distribution. When combined with the results of~\cite{bobkov2011entropy} our results provide additional justification for such approaches when the  posterior is either log-concave or included in subclasses of unimodal distributions that will be discussed in this paper.

The remainder of this paper is organized as follows. %In Section~\ref{sec:motive}, we highlight the motivation of this work and discuss our contributions. 
Section~\ref{sec:int-main} provides  precise statements of our main results, including bounds on the variance of symmetric unimodal mixture densities (Theorem~\ref{thm:mix1_1}, Corollary~\ref{cor:thm1} and Corollary~\ref{thm:mix2_1}), and  bounds on the variance of Lipschitz-continuous unimodal densities with bounded support (Theorem~\ref{thm:combined} and Theorem~\ref{thm:asym}).
In Section~\ref{sec:thm_proof}, we  prove the upper bounds on variance of symmetric unimodal mixture densities and discuss the tightness of the bounds. To prove these variance upper bounds, we assume that the ratio between the maximum and minimum variances of the mixture components is bounded. When this assumption is violated, the variance of symmetric unimodal densities are not necessarily dominated by a monotonic function of entropy power. We demonstrate this by providing a counterexample in Section~\ref{sec:nec}.
%demonstrate a necessary condition for the variance of symmetric unimodal mixture densities to be dominated by a monotonic function of entropy power.
In Section~\ref{sec:Lipschitz}, we prove the upper bounds on variance of Lipschitz-continuous unimodal densities with bounded support. 
More technical aspects of  these proofs are provided in the appendices.
We conclude in Section~\ref{sec:con}.

\medskip
{\it Notation:} We use the $O(\cdot)$ and $\Theta(\cdot)$ notations to describe the asymptotics of real sequences $\{a_n\}$ and $\{b_n\}$: $a_n=O(b_n)$ implies that $a_n\leq M b_n$ for some positive real number $M$ for all $n\geq n_0$; $a_n=\Theta(b_n)$ implies that $a_n\leq M b_n$ and $a_n \geq M' b_n$ for some positive real numbers $M$ and $M'$ for all $n\geq n_0'$.

\section{Statement of Main Results}\label{sec:int-main}
\subsection{Bounds on Variance of Symmetric Unimodal Mixture Densities}\label{sec:int-main_subA}
We first consider symmetric unimodal linear mixture densities of the form
\beq\label{eqn:mixture_ref}
p(x)=\sum_{i=1}^n \alpha_i p_i(x)\quad\text{for}\;\; \alpha_i>0,\;\;\sum_{i=1}^n \alpha_i=1,
\eeq
that is a mixture of exponentially decreasing distributions, $p_i(x)\propto e^{-\beta_i|x-m|^{\theta_i}}$ for any $\beta_i>0$, $\theta_i>0$, $i=1,\dots,n$.

Theorem~\ref{thm:mix1_1} establishes an upper bound on the variance of a linear mixture of exponentially decreasing densities under the assumption that the ratio between  the maximum and minimum variances of the mixture components $p_i(x)$ is bounded.
\begin{thm}\label{thm:mix1_1} {\it Let $p(x)$ be a symmetric unimodal linear mixture density of the form~\eqref{eqn:mixture_ref} with the mixture component $p_i(x)= \frac{1}{Z_i(\theta_i,\beta_i)} e^{-\beta_i|x-m|^{\theta_i}}
$ where $Z_i(\theta_i,\beta_i)$ is a normalizing constant and $\theta_i,\beta_i>0$. Let $\sigma^2_i$ denote the variance of $p_i(x)$, and assume that the ratio of component variances $\sigma^2_i/\sigma^2_j$ is bounded for all $i\neq j$ by some positive constant, i.e., let $r:=\max_{i,j\in\{1,\dots,n\}}\left\{\frac{\sigma_i^2}{\sigma_j^2}\right\}$. Then the variance of the density $p(x)$ satisfies
\beq\label{eqn:thmlwupbds}
\frac{e^{2h(p)}}{2\pi e} \leq  {\text{Var}}(X) \leq B(\pmb{\theta},r) e^{2h(p)}.
\eeq
Here
\beq\label{eqn:defB}
B(\pmb\theta,r)=M(r)\cdot \prod_{i=1}^n\left( \frac{1}{A(\theta_i)}\right)^{\alpha_i},
\eeq
for $\pmb\theta=(\theta_1,\dots,\theta_n)^T$, where
\beq\label{eqn:Atheta1}
A(\theta)=4\theta^{-2}\frac{\left(\Gamma\left({1}/{\theta}\right)\right)^3}{\Gamma\left({3}/{\theta}\right)}e^{{2}/{\theta}},
\eeq
with the Gamma function $\Gamma(t):=\int_0^\infty x^{t-1} e^{-x} dx$ defined for $t>0$, and, 
\beq\label{eqn:M(h)}
M(r):= \frac{(r-1) r^{\frac{1}{r-1}}}{e\log r},\quad r\geq 1.
\eeq
}
\end{thm}

This theorem is proved in Section~\ref{sec:subsec1}.
Equality in the variance lower bound in~\eqref{eqn:thmlwupbds}  is achieved if and only  if $p(x)$ is a Gaussian distribution.
The equality in the variance upper bound  is met when all $p_i(x)$'s are the same distribution, i.e., $p(x)=p_i(x)$ for $\forall i$. 
Therefore, the upper  and lower bounds become equivalent when all $p_i(x)$'s are the same Gaussian distribution.
%More discussion on the tightness of the bounds is provided in Section~\ref{sec:subsec3}
For a mixture distribution of the form~\eqref{eqn:mixture_ref}, possibly with $p_i(x)\neq p_j(x)$ for some pairs of $(i,j)$, the constant scale $B(\pmb\theta,r)$ in \eqref{eqn:defB} is greater than or equal to  $1/2\pi e$ since the geometric mean of $1/A(\theta_i)\geq 1/(2\pi e)$,  $\forall\theta_i$, with weights $\{\alpha_i\}$, is given by
$\prod_{i=1}^n \left(1/A(\theta_i)\right)^{\alpha_i}\geq 1/(2\pi e)$, and $M(r)\geq 1$ with $\lim_{r\to 1}M(r)=1$.

\begin{figure}[t]
\centerline{\includegraphics[scale=0.5]{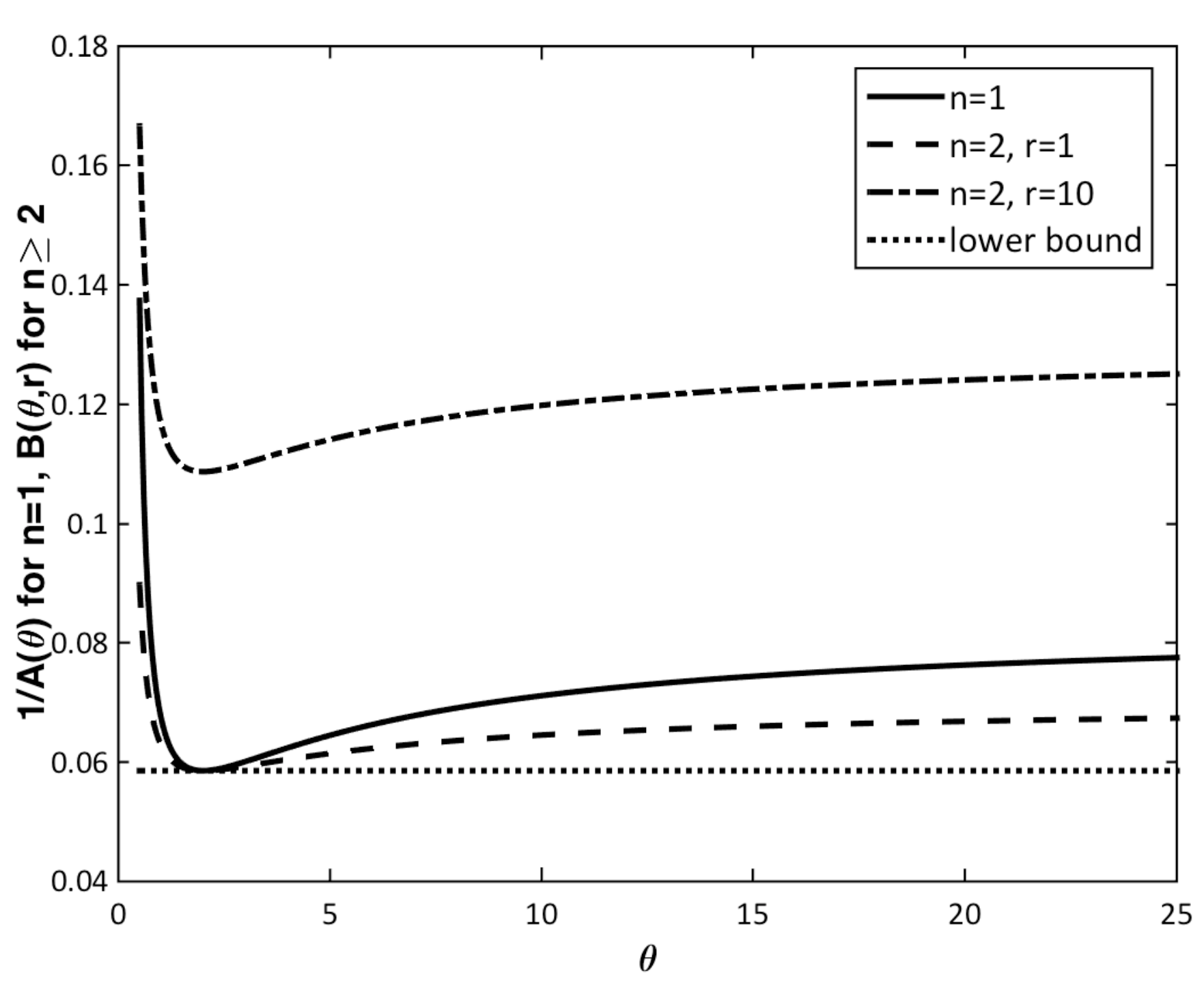}}
\caption{A plot of $1/A(\theta)$ in \eqref{eqn:Atheta1} vs. $\theta$ for $n=1$ (solid line) and plots of $B(\pmb\theta,r)$ in \eqref{eqn:defB} for $n=2$ with mixture weights $\alpha_1=\alpha_2=0.5$ and decaying orders $\theta_1=2$ and $\theta_2=\theta$ when the ratio between variances $r=1$ (dashed line) and when $r=10$ (dash-dot line). The  lower bound $1/2\pi e$ is also shown (dotted line).}
\label{fig:Avstheta}
\end{figure}

In Fig.~\ref{fig:Avstheta}, we compare  the constant factor  $B(\pmb\theta,r)$ in the upper bound for three different mixture distributions. The constant in the lower bound, $1/2\pi e$, is also plotted as the  dotted line. When $n=1$, or equivalently, when all $p_i(x)$'s are the same distribution with $\theta=\theta_i$, $\forall i$, the factor $B(\pmb\theta,r)=1/A(\theta)$. For this case, the equality in the variance upper bound~\eqref{eqn:thmlwupbds} is achieved. The curve $1/A(\theta)$ is plotted as the solid line in Fig.~\ref{fig:Avstheta}. The minimum of $1/A(\theta)$ is achieved at $\theta=2$ (for Gaussian distributions) with the value $1/(2\pi e)\approx 0.0585$. As $\theta$ decreases below $2$, $1/A(\theta)$  increases, and it diverges as $\theta\to 0$. On the other hand, as $\theta$ increases above $2$, $1/A(\theta)$  increases and converges to $1/12\approx 0.0833$ as $\theta\to\infty$. These properties are proven in Lemma~\ref{lem:Atheta}.
We compare this $n=1$ case with those of two different mixture distributions when $n=2$ and the mixture weights $\alpha_1=\alpha_2=0.5$. 
Specifically, consider the case when a Gaussian distribution with $\theta_1=2$ is mixed with another distribution with $\theta_2=\theta$. If the variances of the two distributions are the same, i.e., $r=1$, the resulting $B(\pmb\theta,r)$ is the geometric mean of $1/A(2)=1/(2\pi e)$ and $1/A(\theta)\geq 1/(2\pi e)$, and thus $B(\pmb\theta,r)$ is smaller than $1/A(\theta)$ for all $\theta\neq 2$ and equal to $1/A(\theta)$ at $\theta=2$, as shown by the dashed line.
On the other hand, when the ratio between the variances of two mixture components is $r=10$, $B(\pmb\theta,r)$ is increased by a factor of $M(10)\approx 1.86$ compared to the case when $r=1$, as shown by the dash-dot curve. 

We summarize the properties of $1/A(\theta)$ below.
\begin{lem}\label{lem:Atheta}
{\it
The function $1/A(\theta)$ with $A(\theta)$ in~\eqref{eqn:Atheta1}   is decreasing  on $[0,2]$ and increasing on $[2,+\infty)$. The minimum of $1/A(\theta)$ occurs at $\theta=2$ with the value $1/A(2)=1/(2\pi e)$. As $\theta\to 0$, $\lim_{\theta\to 0} 1/A(\theta)= \infty$, and as $\theta\to\infty$, $\lim_{\theta\to \infty} 1/A(\theta)= 1/12$.  }
\end{lem}
\begin{IEEEproof}
Appendix~\ref{app:lem:Atheta}.
\end{IEEEproof}
By using this lemma, we can further simplify the upper bound on variance of the mixture distribution when all $\theta_i$'s of the mixture component $p_i(x)=\frac{1}{Z_i(\theta_i,\beta_i)} e^{-\beta_i|x-m|^{\theta_i}}$, $i=1,\dots, n$, are lower bounded by some positive constant.
%When all the $\theta_i$'s of the mixture component $p_i(x)=\frac{1}{Z_i(\theta_i,\beta_i)} e^{-\beta_i|x-m|^{\theta_i}}$, $i=1,\dots, n$, are lower bounded by some positive constant, the upper bound on the variance can be further simplified. %When all $\theta_i$'s of $p_i(x)$ are lower bounded by some positive constant, we can simplify the upper bound on variance in Theorem~\ref{thm:mix1_1} only in terms of $M(r)$, as stated in Corollary~\ref{cor:thm1}.
For example, when $\theta_i\geq 1$ for all $i$, the maximum ratio between the upper bound and the lower bound is equal to $\frac{2\pi e}{12}M(r)$ since $\prod_{i=1}^n \left(1/A(\theta_i)\right)^{\alpha_i}\leq 1/12$. When $\theta_i\geq 1/2$ for all $i$, since $\prod_{i=1}^n \left(1/A(\theta_i)\right)^{\alpha_i}\leq (2e^4)/15$, the maximum ratio between the upper bound and the lower bound becomes $\frac{4\pi e^5}{15}M(r) $. Note that when $\theta_i\geq 1/2$, $\forall i$, some of the mixture components $p_i(x)$ can be heavy-tailed distributions, but the variance of $p(x)$ can still be bounded above by a monotone function of the entropy power $e^{2h(p)}$. 

\begin{cor}\label{cor:thm1}{\it
Consider symmetric unimodal densities of the form $p(x)=\sum_{i=1}^n \alpha_i p_i(x)$ for $\alpha_i>0$, $\sum_{i=1}^n \alpha_i=1$, where  $p_i(x)= \frac{1}{Z_i(\theta_i,\beta_i)} e^{-\beta_i|x-m|^{\theta_i}}
$ with a normalizing constant $Z_i(\theta_i,\beta_i)$ and $\beta_i>0$.
When the order $\theta_i\geq 1$ for all $i$, the variance is bounded as
\beq\label{eqn:upvar_subg}
\frac{e^{2h(p)}}{2\pi e} \leq  {\text{Var}}(X) \leq\frac{M(r)}{12} e^{2h(p)}.
\eeq
When $\theta_i\geq 1/2$ for all $i$,
\beq\label{eqn:upvar_heavy}
\frac{e^{2h(p)}}{2\pi e} \leq  {\text{Var}}(X) \leq\frac{2e^4M(r)}{15}  e^{2h(p)},
\eeq
for $M(r)$ in~\eqref{eqn:M(h)}.}
\end{cor}

We next show that Theorem~\ref{thm:mix1_1} implies a similar variance upper bound for a linear mixture of uniform distributions or a mixture of exponentially decreasing distributions and uniform distributions. Let $p$ be the uniform density on the interval $[m-\frac{1}{2\epsilon},m+\frac{1}{2\epsilon}]$ with mean $m\in\mathbb{R}$ and $\epsilon>0$.
When we define a sequence of densities $p_n(x)=e^{-\beta_n |x-m|^n}/Z_n$ with $\beta_n=(2\epsilon)^n$ and the normalizing constant $Z_n$, the sequence of densities $\{p_n\}$ is almost everywhere convergent to $p$ and is dominated by some integrable function g in the sense that $|p_n(x)|\leq g(x)$. By the dominated convergence theorem, the variance and the entropy power of $p_n$ converge to the variance and the entropy power of $p$, respectively.
For the sequence of densities $\{p_n\}$, the exponentially-decreasing order $\theta(=n)$ of the density goes to $+\infty$ as $n\to\infty$. In Lemma~\ref{lem:Atheta}, we showed that $\lim_{\theta\to\infty}1/A(\theta)=1/12$. By using this fact and Theorem~\ref{thm:mix1_1}, we can establish the bound on variance of a linear mixture of uniform distributions.
%We next establish the bound on variance of a linear mixture of uniform distributions with bounded support.
\begin{cor}\label{thm:mix2_1} {\it Assume that $p(x)$ is a bounded-support symmetric unimodal density of the form $p(x)=\sum_{i=1}^n \alpha_{i} p_i(x)$ where $p_i(x)=\mathrm{unif}\left(m-\frac{1}{2\epsilon_i}, m+\frac{1}{2\epsilon_i}\right)$ for $\epsilon_i>0$. Also, assume that $r:=\max_{i,j\in\{1,\dots,n\}}\left\{\frac{\epsilon_i^2}{\epsilon_j^2}\right\}$ is bounded. Then the variance is bounded as
\beq\label{bd:unif}
\frac{e^{2h(p)}}{2\pi e} \leq  {\text{Var}}(X) \leq \frac{M(r) }{12} e^{2h(p)}, 
\eeq
with $M(r)$ in~\eqref{eqn:M(h)}.
}
\end{cor}
%This theorem is proved in Section~\ref{sec:subsec2}. 

The equality in the upper bound on variance is achievable when $\epsilon_i=\epsilon_j$ for $\forall i\neq j$, i.e., when $r=1$, since $\lim_{r\to 1}M(r)=1$. The ratio between the upper and the lower bound, which is proportional to $M(r)$, increases in $r$ and as $r\to \infty$, $M(r)\sim\frac{r}{e\log r}$, which diverges.

%\medskip
\medskip
\begin{rem}\label{rem:scalingfree} The upper bounds~\eqref{eqn:thmlwupbds},~\eqref{eqn:upvar_subg},~\eqref{eqn:upvar_heavy} and~\eqref{bd:unif} on variance of linear mixture densities are scaling-invariant. 
For a linear mixture density $p(x)=\sum_{i=1}^n \alpha_i p_i(x)$, consider a scaled version of the density
$p_\gamma(x)=\gamma p(\gamma x)= \sum_{i=1}^n \alpha_i(\gamma p_i(\gamma x))$ for some $\gamma>0$. In order to show that the upper bounds~\eqref{eqn:thmlwupbds},~\eqref{eqn:upvar_subg},~\eqref{eqn:upvar_heavy} are scaling-invariant, it is enough to show that the ratio $r$ between the maximum and minimum variances of the mixture components $\gamma p_i(\gamma x)$ does not change in $\gamma$. This is true since the variances of all the mixture components $\gamma p_i(\gamma x)$ are scaled by the same constant $1/\gamma^2$.  
\end{rem}
\medskip

\medskip
\begin{rem}
Note that in both Theorem~\ref{thm:mix1_1} and Corollary~\ref{thm:mix2_1}, we assumed the boundedness of the ratio $r$ between the maximum and minimum variances of the mixture components, i.e., $\max_{i,j\in\{1,\dots,n\}}\left\{\sigma_i^2/\sigma_j^2\right\}$ in Theorem~\ref{thm:mix1_1} and $\max_{i,j\in\{1,\dots,n\}}\left\{\epsilon_i^2/\epsilon_j^2\right\}$  in Corollary~\ref{thm:mix2_1} are bounded. 
 In Section~\ref{sec:nec}, by constructing a counterexample, we show that when the ratio between the maximum and minimum variances of the mixture components is not bounded, the variance of a symmetric unimodal mixture density is not necessarily dominated by a monotonic function of the entropy power.
\end{rem}
\medskip

\medskip
\begin{rem}
In Theorem~\ref{thm:mix1_1}, we considered a linear mixture of only exponentially decreasing distributions, $p_i(x)\propto e^{-\beta_i|x-m|^{\theta_i}}$ for any $\beta_i>0$, $\theta_i>0$, $i=1,\dots,n$, and in Corollary~\ref{thm:mix2_1},  we considered a linear mixture of only uniform distributions $p_i(x)=\mathrm{unif}\left(-\frac{1}{2\epsilon_i}+m,\frac{1}{2\epsilon_i}+m\right)$ for any $\epsilon_i>0$, $i=1,\dots,n$. But a similar upper bound holds for any linear mixture of both exponentially decreasing distributions and uniform distributions, since the uniform distributions are the limit of particular exponential distributions. 
\end{rem}
\medskip

\subsection{Bounds on Variance of Lipschitz-Continuous Unimodal Densities with Bounded Support}

We next consider Lipschitz-continuous unimodal densities with bounded support. We first focus our discussion on Lipschitz-continuous  symmetric unimodal densities and establish a variance upper bound in terms of entropy power. We then generalize this result for asymmetric unimodal densities.

With sufficiently large number of mixture components, any symmetric unimodal density can be arbitrarily closely approximated as a linear mixture of  the form~\eqref{eqn:mixture_ref}. This was established for the special case of uniform densities $p_i(x)$ by Feller in~\cite{feller2008introduction}.  
We generalize this result as follows.
Let $p(x)$ be a symmetric unimodal density with bounded support $[m-s,m+s]$. 
%Even though we consider the density with mean $0$, this result can be easily generalized to any symmetric unimodal density with an arbitrary mean $m$.
Suppose that $p(x)$ satisfies the Lipschitz-continuity condition with constant $c_s>0$, i.e.,
\beq\label{eqn:Lipschitz_cl1}
|p(x+y)-p(x)| \leq c_s |y|
\eeq
for any $x, y\in[m-s,m+s]$. 
For such a $p(x)$, we construct a linear mixture density $\bar{p}_n(x)$ that approximates $p(x)$ such that  the difference between entropy powers of $p(x)$ and $\bar{p}_n(x)$ and variances of $p(x)$ and $\bar{p}_n(x)$ can be bounded as
\beq
\begin{split}\label{eqn:apprx_error1}
&e^{2h(\bar{p}_n)}\leq e^{2h(p)}(1+c_1n^{-1}\log n),\\
 &\Big|\int_{m-s}^{m+s} (x-m)^2(p(x)-p_n(x)) dx\Big|\leq c_2n^{-1}
\end{split}
\eeq
for some constants $c_1,c_2>0$. We provide a construction of such a mixture density $\bar{p}_n(x)$ and prove these bounds in Lemma~\ref{lem:existence}  of Section~\ref{sec:Lipschitz_sym}.

We also establish an upper bound on the variance of such a linear mixture density $\bar{p}_n(x)$, denoted $\text{Var}(\bar p_n) $,  such that
\beq\label{eqn:lem3}
\text{Var}(\bar p_n) \leq \frac{c_s s^2 e^{c_s s^2}}{24} e^{2h(\bar p_n)} \left( 1 + c_3 n^{-1} \right)
\eeq
for some constant $c_3 >0$ in Lemma~\ref{lem:bd_linear_mix_new} of Section~\ref{sec:Lipschitz_sym}.

 By combining~\eqref{eqn:apprx_error1} with \eqref{eqn:lem3} and letting $n\to\infty$, we establish an upper bound on variance of any Lipschitz-continuous  symmetric unimodal densities with bounded support,  in terms of the entropy power.

\begin{thm}\label{thm:combined}
{\it
For any  symmetric unimodal density $p(x)$ with bounded support $[m-s,m+s]$, when $p(x)$ satisfies the Lipschitz condition in~\eqref{eqn:Lipschitz_cl1} with constant $c_s>0$, the variance $\var(X)$ can be bounded above and below by a constant scaling of entropy power  as
\beq\label{eqn:approx_1/n}
\frac{e^{2h(p)}}{2\pi e}\leq  {\text{Var}}(X) \leq \frac{c_s s^2 e^{c_s s^2}}{24} e^{2h( p)}.
\eeq
}
\end{thm}

\begin{figure}[t]
\centerline{\includegraphics[scale=0.4]{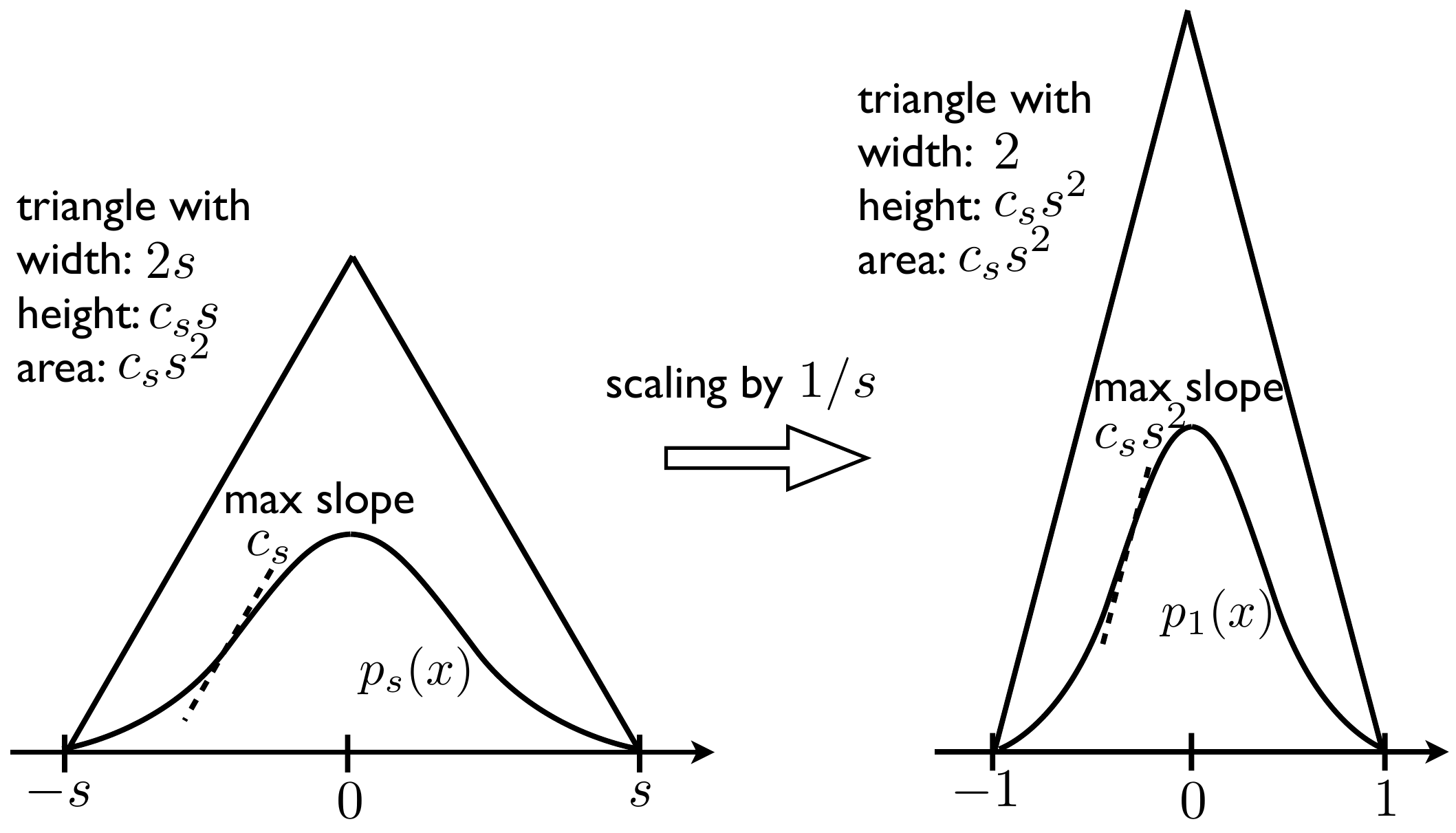}}
\caption{When a symmetric unimodal density $p_s(x)$ with bounded support $[-s,+s]$ satisfies the Lipschitz condition~\eqref{eqn:Lipschitz_cl1} with Lipschitz constant $c_s$, then the density $p_s(x)$ lies below a triangle that is symmetric about $x=0$ with base $2s$ and with slope $c_s$. The area of this triangle $c_ss^2$ is a parameter that determines the variance upper bound in~\eqref{eqn:approx_1/n}. For a density $p_1(x)=sp_s(sx)$, the area of the associated triangle is still $c_ss^2$ since the Lipschitz constant for $p_1$ is $c_ss^2$. Therefore, the parameter $c_ss^2$, which appears in the variance upper bound in~\eqref{eqn:approx_1/n}, is scaling invariant for any $p\equiv p_s$ with a chosen $s$. }
%This parameter is scaling-invariant since the Lipschitz constant $c_s$ is proportional to $s^{-2}$.  }
\label{fig:tri}
\end{figure}

\medskip
\begin{rem}
The tightness of the upper bound in~\eqref{eqn:approx_1/n} depends on the parameter $c_s s^2$ where $c_s$ is the Lipschitz constant and $s$ is a half of the support size.
From the assumption of Lipschitz continuity in~\eqref{eqn:Lipschitz_cl1}, the maximum slope of $p(x)$ is $c_s$ over the support $[m-s,m+s]$. Consider a triangle that is symmetric about  $x=m$ with base $2s$ and with slope on each side  $c_s$, whose area is $c_s s^2$.
When the density $p(x)$ is Lipshitz continuous with the maximum slope less than or equal to $c_s$, then $p(x)$ lies below this symmetric triangle of area $c_ss^2$.
Since the area under $p(x)$ is 1, we can see that $c_ss^2\geq 1$. The area of such a symmetric triangle determines the tightness of the upper bound  in~\eqref{eqn:approx_1/n}.
\end{rem}
\medskip
\begin{rem}\label{rem:lip_sym_scaling}
We also show that the upper bound in~\eqref{eqn:approx_1/n}, or more precisely the parameter $c_s s^2$, is scaling-invariant.
Suppose that $p_s$ is a symmetric unimodal probability density with bounded support $[-s,s]$. We introduce a density $p_1$ defined by
\beq
p_1(x) = s p_s (sx).
\eeq
It is easy to see that $p_1$ is a symmetric unimodal probability density with bounded support $[-1, 1]$. If $p_s$ is Lipschitz continuous with constant $c_s$ as in~\eqref{eqn:Lipschitz_cl1} for any $x,y$, then
\beq
|p_1(x+y)-p_1(x)| = |s p_s(sx+sy) - s p_s (sx)| \leq c_s s^2 |y|.
\eeq
Thus, $p_1$ is Lipschitz continuous with constant $c_1 = c_s s^2$. 
Since it is natural that the Lipschitz constant $c_s$ is proportional to $s^{-2}$, the parameter $c_ss^2$ is independent of the scaling parameter $s$. This fact is illustrated in Fig.~\ref{fig:tri}.
%For the class of distributions $p_s$ that are scaling of $p_1$ with a scaling parameter $s$, it is natural that the Lipschitz constant $c_s$ is proportional to $s^{-2}$. The parameter $c_ss^2$ is thus independent of $s$ and is fixed within the class of distributions. This fact is illustrated in Fig.~\ref{fig:tri}.
Therefore, we can write the upper bound in~\eqref{eqn:approx_1/n} for any $p\equiv p_s$ with a chosen $s$ as
\beq\label{eqn:approx_1/n1}
\frac{e^{2h(p)}}{2\pi e}\leq  {\text{Var}}(X) \leq \frac{c_1 e^{c_1}}{24} e^{2h( p)},
\eeq
with a constant $c_1=c_ss^2>0$, which is scaling-invariant.  
\end{rem}
\medskip

We next generalize this variance upper bound to the case of Lipschitz-continuous asymmetric unimodal densities. 
Consider a unimodal density $p(x)$ with bounded support $[b-s_l,b+s_r]$ with $s_l,s_r>0$. Define $s:=\max\{s_l,s_r\}$. Assume that the unique mode of this density occurs at $x=b$. Let $m$ denote the mean of this density, $m:=\int_{-\infty}^{+\infty} x p(x) dx$. Suppose that $p(x)$ is Lipschitz continuous, i.e.,
\beq\label{eqn:lip_asym}
|p(x+y)-p(x)|\leq c_s|y|
\eeq
for any $x,y\in[b-s_l,b+s_r]$. Then, the variance of this density $p(x)$ can be bounded above by a constant scaling of entropy power minus $(m-b)^2$, where the constant is a function of the Lipschitz constant $c_s$ and the maximum size $s:=\max\{s_l,s_r\}$ of one-sided support from the mode $x=b$. 
%In the following theorem, we provide a variance upper bound for  Lipschitz-continuous asymmetric unimodal densities $p(x)$ with bounded support. 
\begin{thm}\label{thm:asym}
{\it
For any unimodal density $p(x)$ over bounded support $[b-s_l,b+s_r]$ with mode $x=b$ and mean $m$, when $p(x)$ satisfies the Lipschitz condition in~\eqref{eqn:lip_asym}, the variance $\var(X)$ is bounded below and above in terms of the entropy power as
\begin{equation}\label{eqn:var_bd_asym}
\begin{split}
\frac{e^{2h(p)}}{2\pi e}\leq&  \var(X)\leq \frac{c_s s^2 e^{c_s s^2}M\left(128(c_s s^2)^4\right)}{6} e^{2h(p)}-(m-b)^2
\end{split}
\end{equation}
with $s:=\max\{s_l,s_r\}$ and $M(r)$ in~\eqref{eqn:M(h)}.
}
\end{thm}

This theorem is proved in Section~\ref{sec:Lipschitz_asym}.
\medskip
\begin{rem} In Section~\ref{sec:Lipschitz_asym}, we derive an upper bound on the variance of asymmetric unimodal density $p(x)$ in terms of a parameter $r_v\geq 1$ defined for a density $p(x)$ and of $\{\beta_l,\beta_r\}$, which are the areas under the density $p(x)$  in the left and right sides of the mode $x=b$, respectively, i.e.,
\beq\label{eqn:blbrarea}
\beta_l:=\int_{-\infty}^b p(x)dx,\quad\quad \beta_r:=\int_{b}^{+\infty} p(x)dx=1-\beta_l.
\eeq
The resulting upper bound on the variance $\var(X)$ is 
\beq\label{eqn:var_bd_asym_tighter}
\var(X)\leq \frac{c_s s^2 e^{c_s s^2}}{24}\frac{4M(r_v)}{e^{2H_{\sf B}(\beta_l)}}e^{2h(p)} -(m-b)^2.
\eeq
where $H_{\sf B}(\beta_l)=-\beta_l\ln\beta_l-\beta_r\ln\beta_r$. The upper bound in~\eqref{eqn:var_bd_asym} is obtained by showing that $r_v\leq 128\left(c_s s^2\right)^4$ and that $M(r)$ in~\eqref{eqn:M(h)} is an increasing function in $r> 1$ and by using the fact that $H_{\sf B}(\beta_l)\geq 0$ for any $\beta_l>0$.

When the density $p(x)$ is symmetric with mode $x=b$, then the mean $m$ equals the mode $b$ and the parameter $r_v=1$ and the entropy $H_{\sf B}(\beta_l)=\ln2$ since $\beta_l=\beta_r=1/2$. For this case, the upper bound~\eqref{eqn:var_bd_asym_tighter} is specialized to~\eqref{eqn:approx_1/n}, which we derived for symmetric unimodal densities. 

\end{rem}
\medskip

\medskip
\begin{rem} We show that the parameter $c_s s^2$ in the variance upper bound~\eqref{eqn:var_bd_asym} is scaling-invariant due to the similar reasoning as in Remark~\ref{rem:lip_sym_scaling}.  
Suppose that $p_s$ is a Lipschitz-continuous asymmetric unimodal density over bounded support  $[b-s_l,b+s_r]$ with Lipschitz constant $c_s$. Define $s:=\max\{s_l,s_r\}$. Consider a density $p_1$ defined by
\beq
p_1(x)=sp_s(s(x-b)+b).
\eeq 
It can be easily checked that $p_1$ is a Lipschitz-continuous probability density over bounded support $[b-s_l/s,b+s_r/s]$ with mode $x=b$ and Lipschitz constant $c_1:=c_ss^2$. For this density $p_1$, the maximum size of one-sided support from the mode $x=b$ equals $1=\max\{s_l/s,s_r/s\}$. Therefore, for any density $p\equiv p_s$ with a chosen $s$, the variance of $X\sim p$ can be bounded above by 
\beq
 \var(X)\leq \frac{c_1 e^{c_1}M(128(c_1)^4)}{6} e^{2h(p)}-(m-b)^2
\eeq
with a constant $c_1=c_ss^2>0$, which is scaling-invariant. Notice, however, that the term $(m-b)^2$ is proportional to $s^2$. 
\end{rem}
\medskip

\section{Upper Bound on the Variance of Symmetric Unimodal Mixture Densities}\label{sec:thm_proof}
In Section~\ref{sec:subsec0}, we consider generalized Gaussian distributions and show that for this class of distributions, variance and entropy power have an exact monotonic relationship.
By using this result, we provide a proof of Theorem~\ref{thm:mix1_1} in Section~\ref{sec:subsec1}. 

\subsection{Generalized Gaussian Distribution}\label{sec:subsec0}

The unimodality and symmetry of the Gaussian distribution motivate the definition of a larger class of distributions for which variance  has a monotonic relationship to its entropy power.
Consider a symmetric unimodal class of generalized Gaussian densities
\beq\label{eqn:single_unimode}
p(x)=\frac{1}{Z(\theta, \beta)} e^{-\beta|x-m|^\theta},\;\text{ for } \beta,\theta>0
\eeq
where $Z$ is a normalizing constant: $Z(\theta,\beta)=\int_{-\infty}^{\infty}  e^{-\beta|x-m|^\theta}dx$.
The normalizing constant $Z$ and the variance $\text{Var}(X)$ of this distribution can be written in terms of $\beta$ and $\theta$ as follows.
\begin{lem}\label{prop:Z} {\it For distributions of the form $p(x)=\frac{1}{Z(\theta,\beta)} e^{-\beta|x-m|^\theta}$ for $\beta,\theta>0$, the normalizing constant $Z$ and the variance $\text{Var}(X)$ can be expressed as
\begin{align}
Z(\theta, \beta)=&2\beta^{-\frac{1}{\theta}}\theta^{-1} \Gamma\left(\frac{1}{\theta}\right),\label{eqn:Z}\\
\text{Var}(X)=& \beta^{-\frac{2}{\theta}}\frac{\Gamma\left(\frac{3}{\theta}\right)}{\Gamma\left(\frac{1}{\theta}\right)},\label{eqn:sigma}
\end{align}
where  $\Gamma$ denotes the Gamma function $\Gamma(t):=\int_0^\infty x^{t-1} e^{-x} dx$ for $t>0$.  }
\end{lem}
\begin{IEEEproof}%[Proof of Lemma~\ref{prop:Z}]
Appendix~\ref{sec:proof1}
\end{IEEEproof}

We show that for the class of generalized Gaussian distributions in~\eqref{eqn:single_unimode} with a fixed order parameter $\theta$ the variance has an exact monotonic relationship to the entropy power.
Note that this set of symmetric unimodal distributions include heavy-tailed distributions when the order parameter $\theta$ is between $0<\theta<1$. Moreover, this set of distributions is not log-concave for $0<\theta<1$ so the results of~\cite{bobkov2011entropy} do not apply.

\begin{lem}\label{thm:signle_ct} {\it For symmetric unimodal distributions of the form $p(x)=\frac{1}{Z(\theta,\beta)} e^{-\beta|x-m|^\theta}$ for $\beta,\theta>0$, the variance is proportional to the entropy power
\beq\label{eqn:1/Atheta_gGauss}
{\text{Var}}(X) =\frac{1}{A(\theta)} e^{2h(p)},
\eeq
with $A(\theta)$ in~\eqref{eqn:Atheta1}.
}
\end{lem}
\begin{IEEEproof}%[Proof of Proposition~\ref{thm:signle_ct}]
Appendix~\ref{sec:proof2}
\end{IEEEproof}

Important properties of $1/A(\theta)$ are stated in Lemma~\ref{lem:Atheta} of Section~\ref{sec:int-main_subA}. 
By using these properties of $1/A(\theta)$, when $\theta$ is known to be larger than some positive constant, we can bound the variance of the generalized Gaussian distribution $p(x)=\frac{1}{Z(\theta,\beta)} e^{-\beta|x-m|^\theta}$ with $\beta,\theta>0$ in terms of a constant scaling of entropy power, $c e^{2h(p)}$, with $c$ independent of $\theta$.
%By bounding $1/A(\theta)$ in~\eqref{eqn:1/Atheta_gGauss}  for $\theta$ larger than some positive constants, we can find bounds on variance of the generalized Gaussian distributions $p(x)=\frac{1}{Z(\theta,c)} e^{-c|x-m|^\theta}$ with $c>0$. 
\begin{cor}{\it For symmetric unimodal densities of the form $p(x)=\frac{1}{Z(\theta,\beta)} e^{-\beta|x-m|^\theta}$ with $\beta>0$, when $\theta\geq 1$
\beq
{\text{Var}}(X) \leq \frac{1}{12}  e^{2h(p)}.
\eeq When $\theta\geq 1/2$
\beq
{\text{Var}}(X)\leq\frac{2e^4}{15} e^{2h(p)}.
\eeq 
}
\end{cor}

\subsection{Proof of  Theorem~\ref{thm:mix1_1}}\label{sec:subsec1}
We next consider a generalization of Lemma~\ref{thm:signle_ct}  to a much broader class of symmetric unimodal distributions. Let us consider a mixture distribution $p(x)$ composed of a finite number of exponentially decreasing distributions, $p_i(x)=\frac{1}{Z_i(\theta_i,\beta_i)}e^{-\beta_i |x-m|^{\theta_i}}$ with order $\theta_i>0$ and $\beta_i>0$ for $i=1,\cdots, n$, with mixture weights $\alpha_i$, i.e., 
\beq\label{eqn:series}
p(x)=\sum_{i=1}^n \alpha_i \left(\frac{1}{Z_i(\theta_i,\beta_i)} e^{-\beta_i|x-m|^{\theta_i}}\right)
\eeq
where $\alpha_i> 0$ and $\sum_{i=1}^n \alpha_i=1$. All the mixture components $p_i(x)$ have the same mean $m$ to ensure unimodality of $p(x)$. Note that for each mixture component $p_i(x)$, the normalizing constant is $Z_i(\theta_i,\beta_i)=2\beta_i^{-{1}/{\theta_i}}\theta_i^{-1} \Gamma\left({1}/{\theta_i}\right)$ and the variance is $\sigma_i^2:= \beta_i^{-{2}/{\theta_i}}\frac{\Gamma\left({3}/{\theta_i}\right)}{\Gamma\left({1}/{\theta_i}\right)}$, as shown in Proposition~\ref{prop:Z}.
The variance of the symmetric unimodal mixture density~\eqref{eqn:series} is
\beq
\text{Var}(X)=\int_{-\infty}^{\infty} (x-m)^2 p(x) dx=\sum_{i=1}^n \alpha_i \sigma_i^2.
\eeq
For a linear mixture density $p(x)$ of the form~\eqref{eqn:series} we obtain an upper  bound on the variance in terms of the entropy power, as stated in Theorem~\ref{thm:mix1_1}.
Here we present the proof.

Using the concavity of the differential entropy $h(p)$ in distribution $p(x)$,
\beq
h(p)\geq \sum_{i=1}^n \alpha_i h(p_i),
\eeq
and thus
\beq
\begin{split}
e^{2h(p)}\geq& e^{\sum_{i=1}^n 2\alpha_i h(p_i)}=\prod_{i=1}^n\left(e^{2h(p_i)}\right)^{\alpha_i}.
\end{split}
\eeq

As shown in Lemma~\ref{thm:signle_ct}, for $p_i(x)=\frac{1}{Z_i(\theta_i,\beta_i)}e^{-\beta_i |x-m|^{\theta_i}}$,
\beq
\sigma_i^2=\frac{1}{A(\theta_i)} \cdot e^{2h(p_i)},
\eeq
and thus
\beq\label{eqn:split_gm}
e^{2h(p)}\geq {\left(\prod_{j=1}^n {A(\theta_j)}^{\alpha_j}\right)}\cdot \left(\prod_{i=1}^n \left(\sigma_i^2\right)^{\alpha_i}\right).
\eeq

By using the reverse power mean inequality shown in~\cite{specht1960theorie} (English version: p.79 in~\cite{mitrinovic1970analytic}), a lower bound on the geometric mean of $\{\sigma_i^2\}$ with orders $\{\alpha_i\}$ in terms of the arithmetic mean of $\{\sigma_i^2\}$  with orders $\{\alpha_i\}$ is given by
\beq\label{eqn:rev_am_gm_thm2}
\sum_{i=1}^n \alpha_i \sigma_i^2\leq M(r) \prod_{i=1}^n \left(\sigma_i^2\right)^{\alpha_i}
\eeq
where $M(r)$ is defined in \eqref{eqn:M(h)} and $r:=\max_{i,j\in\{1,\dots,n\}}\left\{\frac{\sigma_i^2}{\sigma_j^2}\right\}$.

Since the variance of the mixture distribution $p(x)=\sum_{i=1}^n \alpha_i p_i(x)$ is $\text{Var}(X)=\sum_{i=1}^n \alpha_i \sigma_i^2$,
by combining \eqref{eqn:split_gm} and \eqref{eqn:rev_am_gm_thm2}, we obtain
\beq
\begin{split}
&\text{Var}(X)=\sum_{i=1}^n \alpha_i \sigma_i^2\leq M(r) \prod_{i=1}^n \left(\sigma_i^2\right)^{\alpha_i}\\
&\leq  M(r)\left(\prod_{i=1}^n \left(\frac{1}{A(\theta_i)}\right)^{\alpha_i} \right)e^{2h(p)}. 
\end{split}
\eeq

\section{Mixture Densities with Unbounded Variance Ratio}\label{sec:nec}

In both Theorem~\ref{thm:mix1_1} and Corollary~\ref{thm:mix2_1}, we assumed boundedness of the ratio $r$ between the maximum and minimum variances of the mixture components, i.e., $\max_{i,j\in\{1,\dots,n\}}\left\{\sigma_i^2/\sigma_j^2\right\}$ in Theorem~\ref{thm:mix1_1} and $\max_{i,j\in\{1,\dots,n\}}\left\{\epsilon_i^2/\epsilon_j^2\right\}$  in Corollary~\ref{thm:mix2_1} are bounded. Here we show by a counterexample  that when this assumption is violated, the variance of a symmetric unimodal mixture density is not necessarily dominated by a monotonic function of the entropy power.
%{\textit{necessary}} for existence of an upper bound on variance that is a scaled entropy power of the form $c e^{2h(p)}$ for some constant $c>0$. 
%When the ratio between the variances of the mixture components becomes unbounded, then for the mixture distribution $p(x)$ no such scale factor $c$ exists. 
The following example illustrates this point. Consider a symmetric unimodal distribution composed of two uniform densities of the form
\beq
p(x)=\sum_{i=1}^2 \alpha_i \cdot \mathrm{unif}\left(-\frac{1}{2\epsilon_i},\frac{1}{2\epsilon_i}\right)
\eeq
where $\alpha_i>0$ and $\sum_{i=1}^2 \alpha_i=1$ for $\epsilon_1>\epsilon_2>0$.
The variance of this distribution is equal to
\beq\label{eqn:var_counter}
\text{Var}(X)=\frac{1}{12}\left(\alpha_1\frac{1}{\epsilon_1^2}+\alpha_2\frac{1}{\epsilon_2^2}\right),
\eeq
and the differential entropy of this distribution is
\beq
\begin{split}
h(p)=&-\frac{1}{\epsilon_1}(\alpha_1\epsilon_1+\alpha_2\epsilon_2)\log(\alpha_1\epsilon_1+\alpha_2\epsilon_2)\\
&-\left(\frac{1}{\epsilon_2}-\frac{1}{\epsilon_1}\right)\alpha_2\epsilon_2\log(\alpha_2\epsilon_2).
\end{split}
\eeq
When  $\epsilon_1/\epsilon_2\to\infty$, the limit of the differential entropy becomes
\beq
\lim_{\epsilon_1/\epsilon_2\to\infty} h(p)=-\alpha_1\log\epsilon_1-\alpha_2\log\epsilon_2+H_{\sf B}(\alpha_1)
\eeq
where $H_{\sf B}(\alpha_1)=-\alpha_1\log\alpha_1-(1-\alpha_1)\log(1-\alpha_1)$.
Then, the entropy power becomes
\beq\label{eqn:entpower_counter}
\lim_{\epsilon_1/\epsilon_2\to\infty}e^{2h(p)}=e^{2H_B(\alpha_1)}  \left(\frac{1}{\epsilon_1^2}\right)^{\alpha_1}\left(\frac{1}{\epsilon_2^2}\right)^{\alpha_2}.
\eeq
In order to find an upper bound on the variance in (\ref{eqn:var_counter}) with the entropy power  in \eqref{eqn:entpower_counter}, we need an upper bound on the arithmetic mean $\left(\alpha_1\frac{1}{\epsilon_1^2}+\alpha_2\frac{1}{\epsilon_2^2}\right)$ in terms of the geometric mean $\left(\frac{1}{\epsilon_1^2}\right)^{\alpha_1}\left(\frac{1}{\epsilon_2^2}\right)^{\alpha_2}$. However,  if $\epsilon_2\to 0$ for a fixed $\epsilon_1$, since the arithmetic mean increases on the order of $\Theta\left({1}/{\epsilon_2^2}\right)$ while  the geometric mean on the order of $\Theta\left({1}/{\epsilon_2^{(2\alpha_2)}}\right)$, for $\alpha_2<1$ the variance increases much faster than $e^{2h(p)}$ so that it cannot be bounded above by any constant scaling of entropy power. On the other hand, if $\epsilon_1\to\infty$ for a fixed $\epsilon_2$, the arithmetic mean, which is proportional to the variance, is approximately ${\alpha_2}/\epsilon_2^2$, which is a constant, but the geometric mean, which is proportional to $e^{2h(p)}$, goes to 0 on the order of $\Theta\left({1}/{\epsilon_1^{(2\alpha_1)}}\right)$ for $\alpha_1<1$. Therefore, for both cases satisfying $\epsilon_1/\epsilon_2\to\infty$,  the variance of $p(x)$ cannot be bounded above by a constant scaling of $e^{2h(p)}$. This example shows that when the ratio between variances of mixture components is unbounded, there does not necessarily exist an upper bound on the variance of a symmetric unimodal distribution that is a constant scaling of entropy power.

\section{Upper Bound on Variance of Lipschitz-Continuous Unimodal Density with Bounded Support}\label{sec:Lipschitz}
In Section~\ref{sec:Lipschitz_sym}, we first consider Lipschitz-continuous symmetric unimodal densities and provide a proof of Theorem~\ref{thm:combined}.
In Section~\ref{sec:Lipschitz_asym}, we generalize this result to  Lipschitz-continuous asymmetric unimodal densities and provide a proof of Theorem~\ref{thm:asym}.

\subsection{Proof of Theorem~\ref{thm:combined}: Lipschitz-Continuous Symmetric Unimodal Densities}\label{sec:Lipschitz_sym}
In this section, we show that the variance of  any Lipschitz-continuous symmetric unimodal density $p(x)$ with bounded support can  be bounded above by a scaled entropy power of the form $c e^{2h(p)}$ for some constant $c>0$. We emphasize that here $p(x)$ is not restricted to a mixture density.
To prove the bound, we construct  a linear mixture density $\bar{p}_n(x)$ of uniform densities that approximates $p(x)$. The difference in variances and the difference in entropy powers of $p(x)$ and of $\bar{p}_n(x)$ are shown to be monotonically decreasing as the number of mixture components, $n$, increases. 
Moreover, we establish an upper bound on variance of $X\sim \bar{p}_n(x)$ in terms of the entropy power of $\bar{p}_n(x)$. By combining this result with the fact that $\bar{p}_n(x)$ approximates $p(x)$ with monotonically decreasing differences both in variance and in entropy power, we obtain a bound~\eqref{eqn:approx_1/n} on variance of the Lipschitz-continuous symmetric unimodal density $p(x)$ in terms of entropy power $e^{2h(p)}$, as stated in Theorem~\ref{thm:combined}.

\begin{figure}[t]
\centerline{\includegraphics[scale=0.5]{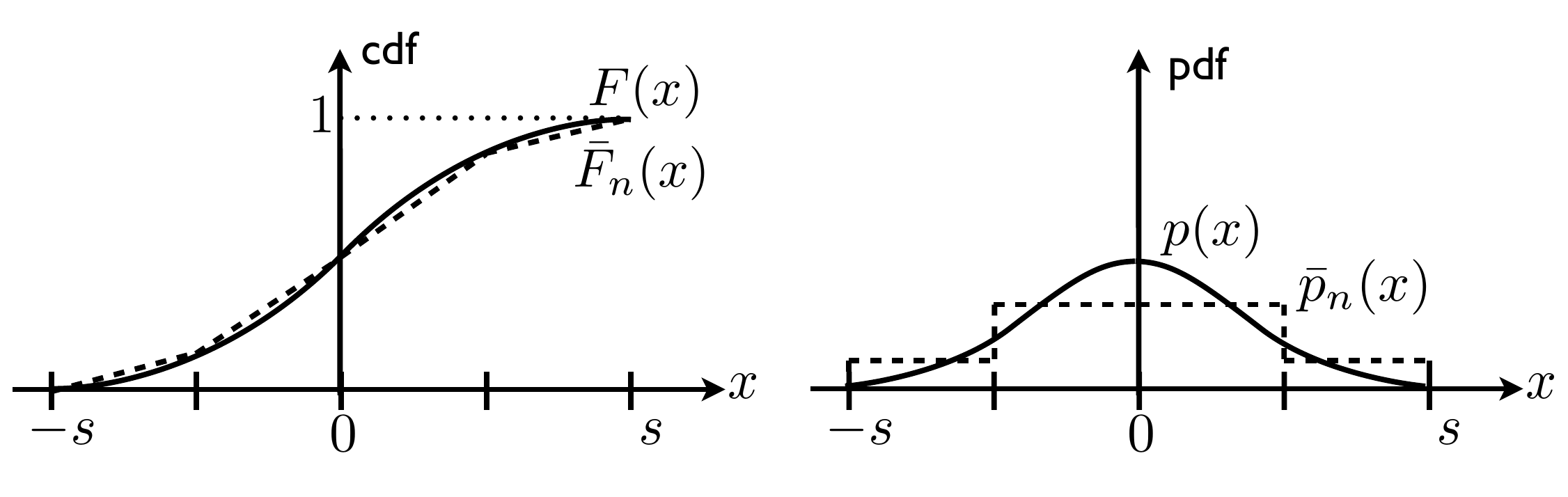}}
\caption{The cumulative distribution function (cdf) $F(x)$ (solid line in the left figure) is approximated by the piecewise linear cdf $\bar{F}_n(x)$ (dashed line in the left figure) such that $\bar{F}_n(x)=F(x)$ at discrete points $x=kh$ for $k\in\{-n,\dots,n\}$ where $nk=s$. The corresponding probability density functions (pdf) $p(x)$ (solid line) and $\bar{p}_n(x)$ (dashed line) are shown in the right figure. The density $\bar{p}_n(x)$ is a linear mixture of uniform densities. }
%This parameter is scaling-invariant since the Lipschitz constant $c_s$ is proportional to $s^{-2}$.  }
\label{fig:cdf}
\end{figure}

Let us first construct a mixture density $\bar{p}_n(x)$  that approximates $p(x)$. Assume that $p(x)$ is a Lipschitz-continuous symmetric unimodal density  with bounded support $[-s,s]$, and denote its cumulative distribution function (cdf) as $F(x)$.
Here for simplicity we consider the density with mean $0$, but this result can be easily generalized to any symmetric unimodal density with an arbitrary mean $m$.
%Even though we consider the density with mean $0$, this result can be easily generalized to any symmetric unimodal density with an arbitrary mean $m$.
Let $\bar{F}_n(x)$ be a cdf having the same value as $F(x)$ at discrete points $x=k h$ for $k\in\{-n,\dots,n\}$ where $nh =s$ and $n$ is an integer. Let $\bar{F}_n(x)$ be linear in each interval $[kh,(k+1)h)$ for all $k\in\{-n,\dots,n-1\}$.
The distribution $F(x)$ is symmetric and  unimodal if and only if $\bar{F}_n(x)$ is symmetric and unimodal for any $h>0$. 
Denote the density of the distribution  $\bar{F}_n(x)$ by $\bar{p}_n(x)$.
Since $\bar{F}_n(x)$ is a piecewise linear distribution, its density $\bar{p}_n(x)$ can be written as a step function, which is a linear mixture of uniform distributions with mean $0$, i.e.,
\beq\label{eqn:pnxalphai}
\bar{p}_n(x)=\sum_{t=1}^n \alpha_t \cdot \mathrm{unif}(-th,t h).
\eeq
Fig.~\ref{fig:cdf} illustrates the piecewise linear cdf $\bar{F}_n(x)$ that approximates $F(x)$ and the corresponding probability density functions (pdfs) $\bar{p}_n(x)$ and $p(x)$.

From the definition of $\bar{F}_n(x)$, $F(x)=\bar{F}_n(x)$ at $x=kh$ for $k\in\{-n,\dots,n\}$, and this implies that
\beq
\begin{split}
&\int_{kh}^{(k+1)h} p(x) dx=\int_{kh}^{(k+1)h} \bar{p}_n(x) dx.
\end{split}
\eeq
Since $\bar{p}_n(x)$ is constant in each interval $[{kh},{(k+1)h})$, the value of $\bar{p}_n(x)$ in this interval equals $\frac{1}{h}\int_{kh}^{(k+1)h} p(x) dx$.
From the fact that $\bar{p}_n(x)$ is equal to the average value $\frac{1}{h}\int_{kh}^{(k+1)h} p(x) dx$ in each interval  $[{kh},{(k+1)h})$ for $k\in\{-n,\dots,n-1\}$ and  that $p(x)$ is monotone in each interval, we know that there exists at least one $x\in[{kh},{(k+1)h})$ for each $k\in\{-n,\dots,n-1\}$ such that $\bar{p}_n(x)=p(x)$.

%The values of $\alpha_t$'s in~\eqref{eqn:pnxalphai} are determined as follows.
From the representation of $\bar{p}_n(x)$ in~\eqref{eqn:pnxalphai}, for any $x\in[{kh},{(k+1)h})$ with $k\in\{-n,\dots,n-1\}$,
\beq
\bar{p}_n(x)=\sum_{t=|k|}^n \alpha_t\frac{1}{2th}=\frac{1}{h}\int_{kh}^{(k+1)h} p(x) dx.
\eeq
These relations determine $\alpha_t$ for $t=1,\dots, n-1$ and $\alpha_n$ as 
\beq\label{eqn:alphai_or}
\begin{split}
\alpha_t=&2t\left(\int_{-th}^{-(t-1)h}p(x) dx-\int_{-(t+1)h}^{-th}p(x) dx \right),\\
=&2t\int_{-(t+1)h}^{-th} \left( p(x+h)-p(x)\right) dx,\\%\;\;\text{for}\;\; t=1,\dots,n-1,\\
\alpha_n=&2n\left(\int_{-s}^{-s+h}p(x) dx\right).
%\\
%= & 2i\left((F(-s+(n-i+1)h)-F(-s+(n-i)h))-(F(-s+(n-i)h)-F(-s+(n-i-1)h))\right),
\end{split}
\eeq
Since $p(x)$ increases  monotonically in $[-s,0]$, we can check that $\alpha_i> 0$ for all $i\in\{1,\dots,n\}$. From the definition of $\bar{p}_n(x)$, $\sum_{i=1}^n \alpha_i=1$.

We show that both the difference in variances and the difference in entropy powers of $p(x)$ and of $\bar{p}_n(x)$ decrease monotonically in $n$ as $n$ goes to infinity. 
\begin{lem}\label{lem:existence}
{\it
When a symmetric unimodal density $p(x)$ with bounded support $[-s,s]$ is Lipschitz continuous, $\bar{p}_n(x)$ defined in~\eqref{eqn:pnxalphai} approximates $p(x)$ with bounded variance difference and bounded entropy power difference, which are monotonically decreasing in $n$, such that
\beq
\begin{split}
&e^{2h(\bar{p}_n)}\leq e^{2h(p)}(1+c_1n^{-1}\log n),\\
 &\Big|\int_{-s}^s x^2p(x) dx-\int_{-s}^s x^2\overline p_n(x) dx\Big|\leq c_2n^{-1},
\end{split}
\eeq
for some constants $c_1,c_2>0$.
}
\end{lem}
\begin{IEEEproof}
Appendix~\ref{sec:proof_lem_sec5}.
\end{IEEEproof}

We also prove that for the linear mixture density $\bar{p}_n(x)$ in~\eqref{eqn:pnxalphai} with $\alpha_t$'s in~\eqref{eqn:alphai_or}, the variance, denoted $\text{Var}(\bar p_n) $, can be upper bounded in a scaled entropy power in the form of $ce^{2h(\bar{p}_n)}$ for some constant $c>0$.
\begin{lem}\label{lem:bd_linear_mix_new}
{\it
For a linear mixture density $\bar{p}_n(x)$ in~\eqref{eqn:pnxalphai} with $\alpha_t$'s in~\eqref{eqn:alphai_or}, the variance can be upper bounded as 
\beq
\text{Var}(\bar p_n) \leq \frac{c_s s^2 e^{c_s s^2}}{24} e^{2h(\bar p_n)} \left( 1 + c_3 n^{-1} \right)
\eeq
for some constant $c_3 > 0$.}
\end{lem}
\begin{IEEEproof}
Appendix~\ref{sec:proof_var_bd_linear_mix_new}
\end{IEEEproof}

By combining Lemma~\ref{lem:existence} and Lemma~\ref{lem:bd_linear_mix_new}, we can prove that for any  Lipschitz-continuous symmetric unimodal density $p(x)$ with bounded support, the variance can be bounded above by a constant scaling of entropy power:
\beq
\var(X)\leq \frac{c_s s^2 e^{c_s s^2}}{24}e^{2h(p)}(1+cn^{-1}\log n),
\eeq
for some constant $c>0$.
Now letting $n\to \infty$,
\beq
\var(X)\leq \frac{c_s s^2 e^{c_s s^2}}{24}e^{2h(p)}.
\eeq
When combined with the lower bound on variance~\eqref{eqn:est_counter_Fano}, this proves Theorem~\ref{thm:combined}.

\subsection{Proof of Theorem~\ref{thm:asym}: Lipschitz-Continuous Asymmetric Unimodal Densities}\label{sec:Lipschitz_asym}
In this section, we establish a variance upper bound for Lipschitz-continuous asymmetric unimodal distributions by using the result of the previous section.
Consider a Lipschitz-continuous unimodal density $p(x)$ over bounded support $[b-s_l,b+s_r]$ with mode $x=b$ and mean $m=\int_{-\infty}^{+\infty}xp(x)dx$.
Let $\{\beta_l,\beta_r\}$ denote the areas under the density $p(x)$  in the left and right sides of the mode $x=b$, respectively, as defined in~\eqref{eqn:blbrarea}.
Assume that the density $p(x)$ satisfies the Lipschitz-continuity condition with constant $c_s>0$, i.e.,
\beq\label{eqn:lip_asym_latersec}
|p(x+y)-p(x)|\leq c_s|y|
\eeq
for any $x,y\in[b-s_l,b+s_r]$.

By using $p(x)$, we define two symmetric unimodal densities $p_l(x)$ and $p_r(x)$ with the unique mode $x=b$ such that
\beq
\begin{split}\label{eqn:plprdef}
&p_l(x)=\begin{cases}
\frac{1}{2\beta_l}p(x),&\quad x\in[b-s_l, b],\\
\frac{1}{2\beta_l}p(-x+2b),&\quad  x\in(b,b+s_l],\\
0, &\quad {\text{otherwise}},
\end{cases}\\
&p_r(x)=\begin{cases}
\frac{1}{2\beta_r}p(-x+2b),&\quad x\in[b-s_r, b]\\
\frac{1}{2\beta_r}p(x),&\quad  x\in(b,b+s_r],\\
0, &\quad {\text{otherwise}}.
\end{cases}
\end{split}
\eeq
Since both $p_l$ and $p_r$ are Lipschitz-continuous symmetric unimodal densities with Lipschitz constant $c_s>0$ and support $[b-s_l,b+s_l]$ and $[b-s_r,b+s_r]$, respectively, by using Theorem~\ref{thm:combined}, the variances of $X\sim p_l$ and of $X\sim p_r$ are bounded above in terms of their respective entropy powers as
\beq
\begin{split}\label{eqn:var_up_pl_pr}
\var(p_l)\leq &  \frac{c_s s_l^2 e^{c_s s_l^2}}{24}e^{2h(p_l)},\\
\var(p_r)\leq &  \frac{c_s s_r^2 e^{c_s s_r^2}}{24}e^{2h(p_r)}.
\end{split}
\eeq
Moreover, the variance $\var(X)$ of the original asymmetric unimodal density $p(x)$ can be represented as
\beq\label{eqn:var_exp_pl_pr}
\var(X)=\beta_l\var(p_l)+\beta_r\var(p_r)-(m-b)^2,
\eeq
and the entropy power $e^{h(p)}$ of $p(x)$ as
\beq\label{eqn:ent_exp_pl_pr}
e^{2h(p)}= \frac{1}{4}\left(e^{2h(p_l)}\right)^{\beta_l}\left(e^{2h(p_r)}\right)^{\beta_r}e^{2H_{\sf B}(\beta_l)}
\eeq
where $H_{\sf B}(\beta_l)=-\beta_l\ln \beta_l-\beta_r \ln \beta_r$.

From the reverse power mean inequality~\cite{specht1960theorie} (English version: p.79 in~\cite{mitrinovic1970analytic}), we have
\beq
\beta_l\var(p_l)+\beta_r\var(p_r)\leq M(r_v) \left(\var(p_l)\right)^{\beta_l}\left(\var(p_r)\right)^{\beta_r}
\eeq
for $r_v:=\max\{\var(p_l)/\var(p_r),\var(p_r)/\var(p_l)\}$ with $M(r)$ in~\eqref{eqn:M(h)}. By applying this inequality to~\eqref{eqn:var_exp_pl_pr}, we obtain
\beq
\var(X)\leq M(r_v)\left(\var(p_l)\right)^{\beta_l}\left(\var(p_r)\right)^{\beta_r}-(m-b)^2. 
\eeq
Moreover, by using the variance upper bounds in~\eqref{eqn:var_up_pl_pr} and the fact that $\beta_l+\beta_r=1$, we obtain
\beq
\var(X)\leq \frac{c_s s^2 e^{c_s s^2}}{24}M(r_v)e^{2\beta_l h(p_l)}e^{2\beta_r h(p_r)}-(m-b)^2
\eeq
for $s:=\max\{s_l,s_r\}$. Finally, by using~\eqref{eqn:ent_exp_pl_pr}, the variance can be bounded above in terms of $e^{2h(p)}$ as
\beq\label{eqn:var_asym_inter}
\var(X)\leq \frac{c_s s^2 e^{c_s s^2}}{24}\frac{4M(r_v)}{e^{2H_{\sf B}(\beta_l)}}e^{2h(p)} -(m-b)^2.
\eeq

In Lemma~\ref{lem:bound_r_v}, we show that  $r_v\leq 128 (c_s s^2)^4$. Furthermore, notice that $M(r)$ is an increasing function of $r$ for $r>1$, which can be easily check from that
\beq
	M'(r) = \frac{r^{\frac{1}{r-1}- 1} (r-\log r -1) (-r+ r \log r + 1)}{e(r-1) \log^2 r} > 0,
\eeq
where we used that $\log r < r-1 < r \log r$ for $r>1$.  By using these results and the fact that $H_{\sf B}(\beta_l)\geq 0$, the variance upper bound in~\eqref{eqn:var_asym_inter} can be further simplified as 
\beq\label{eqn:var_asym_final}
\var(X)\leq \frac{c_s s^2 e^{c_s s^2}M\left(128\left(c_s s^2\right)^4\right)}{6} e^{2h(p)}-(m-b)^2.
\eeq 
This completes the proof of Theorem~\ref{thm:asym}.

\begin{lem}\label{lem:bound_r_v}
{\it
Consider two Lipschitz-continuous symmetric unimodal densities $p_l$ and $p_r$ in~\eqref{eqn:plprdef}, defined in terms of a Lipschitz-continuous unimodal density $p$ with bounded support $[b-s_l,b+s_r]$ having the unique mode at $x=b$. Let $s=\max\{s_l,s_r\}$, and assume that the density $p$ satisfies the Lipschitz-continuity condition~\eqref{eqn:lip_asym_latersec} with constant $c_s$. 
Then, the parameter $r_v:=\max\{\var(p_l)/\var(p_r),\var(p_r)/\var(p_l)\}$ is bounded as
\beq
r_v\leq 128(c_s s^2)^4.
\eeq
}
\end{lem}
\begin{IEEEproof}
Appendix~\ref{app:lem:bound_r_v}.
\end{IEEEproof}

\section{Conclusion and Future Directions}\label{sec:con}
In this paper, we established upper bounds on the variance of subclasses of unimodal distributions in terms of the entropy power. 
We first considered symmetric unimodal mixture densities of exponentially decreasing distributions and uniform distributions. 
%considered two representations of mixture distributions, one a mixture of exponentials and the other a mixture of uniform distributions. 
%There are two possible representations of symmetric unimodal distribution based on two difference types of basis including exponentially decreasing distribution and uniform distribution. 
The tightness of the upper bounds on variance depends on the ratio between the maximum and minimum variances of the mixture components.
%the ratio between the maximum and minimum variances and the exponents of mixture components.
%Therefore, given a symmetric unimodal distribution, we need to choose the right set of basis to approximate the original distribution in order to find the tightest bound on variance in terms of entropy power.
By constructing a counterexample, we showed that when the ratio between variances of the mixture components is unbounded, there may not necessarily exist an upper bound on  the variance of symmetric unimodal mixture densities that is a constant scaling of entropy power. 
We also showed that the variance of any Lipschitz-continuous unimodal density with a bounded support can be upper bounded in terms of the entropy power. All the upper bounds on the variance of  unimodal densities presented in this paper are scaling-invariant. 
%a variance difference monotonically decreasing in the number of mixture components. However, the way to find the best linear mixture approximation of symmetric unimodal distribution that provides the tightest upper bound on variance is unknown; this would be interesting to identify in the future work. 

In signal processing, adaptive sensing, and machine learning,  information theoretic surrogates such as Kullback-Leibler divergence, entropy, and Fisher information have been widely adopted in place of task-specific cost functions such as mean squared error or probability of classification error. 
Since such task-specific cost functions are often intractable, information-theoretic surrogates are used as natural objectives for developing waveforms or sensor selection strategies to collect and/or filter information. The results reported in this paper can be used to justify the use of differential  entropy as a surrogate for the mean squared error in such applications with a provable performance guarantee, when the posterior distribution is Lipschitz-continuous and unimodal with bounded support or when it can be approximated by a linear mixture of exponentially decreasing densities and uniform densities. 
% is symmetric and unimodal such as the generalized Gaussian distributions or can be approximated as a linear mixture of either exponentially decreasing densities or uniform densities.

One of the interesting future research directions is to extend this work to a multivariate setting.
Concentration inequalities for functions of a large number of random variables have been an active research topic with diverse applications in high-dimensional statistics, machine learning and information theory.
Recent progress in concentration inequalities and their applications to information theory can be found in an excellent monograph~\cite{raginsky2014concentration}. 
Among many interesting aspects, we are particularly interested in finding conditions on the multivariate distribution to derive a concentration inequality in terms of entropy power. 
When the random variables composing a $k$-dimensional random vector $\mathbf{X}=[X_1,X_2,\dots,X_k]$ are mutually independent and the marginal distribution $p_i$ of each random variable $X_i$ are unimodal and satisfying conditions discussed in this paper, the variance $\var(X_i)$ of each random variable $X_i$ is upper bounded by $\var(X_i)\leq c_i e^{2h(p_i)}$ for some constant $c_i>0$ and the determinant of the $k\times k$ covariance matrix $\mathbf{\Sigma}$ equals $|\mathbf{\Sigma}|=\prod_{i=1}^k \var(X_i)$. Moreover, the entropy power $e^{2h(\mathbf{X})}$ equals $e^{2h(\mathbf{X})}=\prod_{i=1}^k e^{2h(p_i)}$.
By using these facts, we can easily extend our work to the multivariate setting and show that 
\beq
|\mathbf{\Sigma}|\leq c e^{2h(\mathbf{X})}
\eeq
for some positive constant $c>0$. Therefore, for this case of the multivariate distribution with independent random variables each satisfying the Lipschitz-continuity and unimodality, as the entropy $h(\mathbf{X})$ of the random vector $\mathbf{X}$ tends to $-\infty$, the determinant of the covariance matrix, which is the product of the variances of the marginal distributions, converges to 0 and the random vector concentrates around its expectation $\E[\mathbf{X}]$. One might want to find a similar concentration result for more general multivariate distributions by bounding the determinant of the covariance matrix. But it is important to note that the determinant of the covariance matrix does not always capture the concentration of the $k$-dimensional distribution when the random variables are not mutually independent. For example, when $X_1=X_2=\dots=X_k$, regardless of how large the variance of $X_i$ is, the determinant of the covariance matrix equals 0. Therefore, an interesting open question is to find a proper metric which captures the concentration of a multivariate distribution and to derive a concentration inequality for the metric in terms of entropy power for an appropriate class of distributions.

%\section*{acknowledgments}
%This research was partially supported by US Army Research Laboratory grant  W911NF-11-1-0391.
%This research was supported in part by ARO grants W911NF-11-1-0391 and W911NF-15-1-0479.
\appendices
\section{Proof of Lemma~\ref{lem:Atheta}}\label{app:lem:Atheta}
In Lemma~\ref{lem:Atheta}, we show that the function $1/A(\theta)$ is decreasing  on $[0,2]$ and increasing on $[2,+\infty)$ where
\beq
A(\theta)=4\theta^{-2}\frac{\left(\Gamma\left({1}/{\theta}\right)\right)^3}{\Gamma\left({3}/{\theta}\right)}e^{{2}/{\theta}}.
\eeq
The minimum of $1/A(\theta)$ occurs at $\theta=2$ with the value $1/A(2)=1/(2\pi e)$. As $\theta\to 0$, $\lim_{\theta\to 0} 1/A(0)= \infty$, and as $\theta\to+\infty$, $\lim_{\theta\to +\infty} 1/A(\theta)= 1/12$.  
To prove this lemma, we first show that $A(\theta)$ is unimodal with the maximum reached at 2, i.e., increasing on $[0,2]$ and decreasing on $[2,+\infty)$.
The derivative of $A(\theta)$ with respect to $\theta$ equals
\begin{equation}
\begin{split}
&A'(\theta)\\
&=4\theta^{-2}\frac{\left(\Gamma(1/\theta)\right)^2}{\Gamma\left(3/\theta\right)}e^{2/\theta}\Big(-2\theta^{-1}\Gamma(1/\theta)+\frac{\Gamma(1/\theta)\Gamma'(3/\theta)(3/\theta^2)}{\Gamma(3/\theta)}-3\Gamma'(1/\theta)(1/\theta^2)-\Gamma(1/\theta)(2/\theta^2)\Big)
\end{split}
\end{equation}
where $\Gamma(z)$ is the Gamma function $\Gamma(z):=\int_0^\infty x^{z-1} e^{-x} dx$ with $z>0$ and $\Gamma'(z)$ is the derivative of $\Gamma(z)$, which equals $\Gamma'(z)=\Gamma(z)\psi_0(z)$ where $\psi_0(z)$ is the polygamma function 
\beq\label{eqn:psi_0}
\psi_0(z)=-\left[\frac{1}{z}+\gamma+\sum_{n=1}^{\infty}\left(\frac{1}{n+z}-\frac{1}{n}\right)\right]
\eeq
with the Euler-Mascheroni constant $\gamma$. By using $\Gamma'(z)=\Gamma(z)\psi_0(z)$, $A'(\theta)$ can be written as
\beq
\begin{split}
&A'(\theta)=4\theta^{-2}\frac{\left(\Gamma(1/\theta)\right)^3}{\Gamma\left(3/\theta\right)}e^{2/\theta}\left(-\frac{2}{\theta}\left(1+\frac{1}{\theta}\right)+\frac{3}{\theta^2}\left(\psi_0\left(\frac{3}{\theta}\right)-\psi_0\left(\frac{1}{\theta}\right)\right)\right),
\end{split}
\eeq
and by using~\eqref{eqn:psi_0},  $A'(\theta)$ can be further simplified as
\begin{equation}
\begin{split}
&A'(\theta)=4\theta^{-4}\frac{\left(\Gamma(1/\theta)\right)^3}{\Gamma\left(3/\theta\right)}e^{2/\theta}\left(-2+3\left(\phi\left(\frac{3}{\theta}\right)-\phi\left(\frac{1}{\theta}\right) \right)\right)
\end{split}
\end{equation}
where 
\beq
\phi(x)=\sum_{n=1}^\infty \left(\frac{1}{n}-\frac{1}{n+x}\right),
\eeq
which converges absolutely on $(0,\infty)$.

Since $4\theta^{-4}\frac{\left(\Gamma(1/\theta)\right)^3}{\Gamma\left(3/\theta\right)}e^{2/\theta}>0$ for $\theta>0$, the sign of $A'(\theta)$ is determined by $\left(-2+3\left(\phi\left({3}/{\theta}\right)-\phi\left({1}/{\theta}\right) \right)\right)$.
When we define $f(\theta):=\phi\left({3}/{\theta}\right)-\phi\left({1}/{\theta}\right) $, we can show that $f(\theta)$ is a strictly decreasing function of $\theta$. In particular,
\beq\label{f1}
	f(\theta) > f(2) = \frac{2}{3} \qquad \text{for } \theta < 2,
\eeq
and
\beq\label{f2}
	f(\theta) < f(2) = \frac{2}{3} \qquad \text{for } \theta > 2.
\eeq
For ease of notation, we let $a=1/\theta$ and show that $f(\theta)=f(1/a)=\phi(3a)-\phi(a)$ is a strictly increasing function of $a$.
Note that
\beq
	\phi'(x) = \sum_{n=1}^{\infty} \frac{1}{(n+x)^2}.
\eeq
It suffices to show that $(\phi(3a) - \phi(a))' = 3\phi'(3a) - \phi'(a) > 0$. This can be easily checked from that
\beq \begin{split}
	&\phi'(3a) = \sum_{n=1}^{\infty} \frac{1}{(n+3a)^2} \\
	&= \sum_{m=1}^{\infty} \Big( \frac{1}{(3m-2+3a)^2} + \frac{1}{(3m-1+3a)^2} + \frac{1}{(3m+3a)^2} \Big) \\
	&= \frac{1}{9} \sum_{m=1}^{\infty} \left( \frac{1}{(m+a-\frac{2}{3})^2} + \frac{1}{(m+a-\frac{1}{3})^2} + \frac{1}{(m+a)^2} \right) \\
	&> \frac{1}{9} \sum_{m=1}^{\infty} \left( \frac{1}{(m+a)^2} + \frac{1}{(m+a)^2} + \frac{1}{(m+a)^2} \right) \\
	&= \frac{1}{3} \sum_{m=1}^{\infty} \frac{1}{(m+a)^2} \\
	&= \frac{\phi'(a)}{3}.
\end{split} \eeq
This proves~\eqref{f1} and~\eqref{f2}, which imply that $A'(\theta)>0$ for $0<\theta<2$, $A'(\theta)=0$ at $\theta=2$, and $A'(\theta)<0$ for $\theta>2$. Thus, $A(\theta)$ is increasing on $[0,2]$ and decreasing on $[2,+\infty)$.

We next prove the statement that as $\theta\to 0$, $\lim_{\theta\to 0} 1/A(0)= \infty$. 
By using Stirling's formula, as $z\to \infty$,
$
\Gamma(z+1)\sim \sqrt{2\pi z}\left(\frac{z}{e}\right)^{z}.
$
Using this approximation and the property that $\Gamma(z+1)=z\Gamma(z)$, it can be shown that as $\theta\to 0$,
$
\frac{1}{\theta}\Gamma\left(\frac{1}{\theta}\right)=\Gamma\left(\frac{1}{\theta}+1\right)\sim \sqrt{2\pi\frac{1}{\theta}}\left(\frac{1}{\theta e}\right)^{\frac{1}{\theta}}$ and
$\frac{3}{\theta}\Gamma\left(\frac{3}{\theta}\right)=\Gamma\left(\frac{3}{\theta}+1\right)\sim \sqrt{2\pi\frac{3}{\theta}}\left(\frac{3}{\theta e}\right)^{\frac{3}{\theta}}$.
Therefore, as $\theta\downarrow 0$, i.e., $1/\theta\to \infty$,
\beq
A(\theta)\sim 8\sqrt{3}\pi \theta^{-1} e^{\frac{1}{\theta}(2-3\log3)},
\eeq
which tends to 0 since $(2-3\log3)<0$. Therefore, $1/A(\theta)$ diverges as $\theta\to 0$.
%On the other hand, as $\theta$ increases above $2$, $1/A(\theta)$  increases and converges to $1/12\approx 0.0833$.

We next show that as $\theta\to\infty$, $\lim_{\theta\to \infty} 1/A(\theta)= 1/12$. 
From  the Gamma function property that $z\Gamma(z)=\Gamma(z+1)$ and $\Gamma(1)=1$, it follows that $\lim_{z\to 0} z\Gamma(z)=1$, i.e, $ \Gamma(z) \sim \frac{1}{z}
$ as $z\to 0$.
Thus, when $\theta\to \infty$,
% i.e., $1/\theta\to 0$,
$
A(\theta)\sim 4\theta^{-2}\frac{\theta^3}{\theta/3}\exp\left(2/\theta\right)
$
and $\lim_{\theta\to\infty}1/A(\theta)=1/12$.

\section{Proof of Lemma~\ref{prop:Z}}\label{sec:proof1}
The normalizing constant $Z$ is
$
Z(\theta,\beta)=\int_{-\infty}^{\infty} e^{-\beta|x-m|^\theta}dx=2\int_{0}^{\infty} e^{-\beta x^\theta}dx.
$
Let $\beta x^\theta=y$. Then, $\beta\theta x^{\theta-1}dx=dy$ and $x^{\theta-1}=(y/\beta)^{\frac{\theta-1}{\theta}}$. Thus, $Z$ can be written in terms of $y$ as
\begin{equation}
\begin{split}\label{eqn:Z11} 
Z(\theta,\beta)=&2\int_0^{\infty} \beta^{-\frac{1}{\theta}}\theta^{-1}y^{-1+\frac{1}{\theta}}e^{-y} dy=2 \beta^{-\frac{1}{\theta}}\theta^{-1}\Gamma\left(\frac{1}{\theta}\right).
\end{split}
\end{equation}
In a similar way,
\begin{equation}
\begin{split}\label{eqn:sigma11}
&\text{Var}(X)=\frac{1}{Z(\theta,\beta)}\int_{-\infty}^{\infty} (x-m)^2 e^{-\beta|x-m|^\theta}dx\\
&=\frac{2}{Z(\theta,\beta)}\int_{0}^{\infty} x^2 e^{-\beta x^\theta}dx\\
&=\frac{2}{Z(\theta,\beta)} \beta^{-\frac{3}{\theta}} \theta^{-1}\int_{0}^{\infty} y^{-1+\frac{3}{\theta}} e^{-y} dy\\
&=\frac{2}{Z(\theta,\beta)} \beta^{-\frac{3}{\theta}} \theta^{-1} \Gamma\left(\frac{3}{\theta}\right)
=\beta^{-\frac{2}{\theta}}\frac{\Gamma\left(\frac{3}{\theta}\right)}{\Gamma\left(\frac{1}{\theta}\right)}.
\end{split}
\end{equation}

\section{Proof of Lemma~\ref{thm:signle_ct}}\label{sec:proof2}
For $p(x)=\frac{1}{Z(\theta,\beta)} e^{-\beta|x-m|^\theta}$ where $\beta,\theta>0$, the differential entropy can be directly calculated as follows.
\beq
\begin{split}
h(p)&=-\int_{-\infty}^{\infty} \frac{e^{-\beta|x-m|^\theta}}{Z(\theta,\beta)} \left(\log\frac{1}{Z(\theta,\beta)}-\beta|x-m|^\theta\right)dx\\
&=\log Z(\theta,\beta)+\frac{2\beta}{Z(\theta,\beta)}\int_{0}^{\infty}x^{\theta} e^{-\beta x^\theta}dx.
\end{split}
\eeq
Let $\beta x^\theta=y$. Then, $\beta\theta x^{\theta-1}dx=dy$ and $x=(y/\beta)^{\frac{1}{\theta}}$. The differential entropy becomes
\beq
\begin{split}
h(p)=&\log Z(\theta,\beta)+\frac{2\beta}{Z(\theta,\beta)}\int_{0}^{\infty} \beta^{-1}\theta^{-1} (y/\beta)^{\frac{1}{\theta}}e^{-y} dy\\
=&\log Z(\theta,\beta)+\frac{2\beta^{-\frac{1}{\theta}}\theta^{-1}}{Z(\theta,\beta)} \int_{0}^{\infty} y^{\frac{1}{\theta}}e^{-y}dy\\
=&\log Z(\theta,\beta)+\frac{2\beta^{-\frac{1}{\theta}}\theta^{-1}}{Z(\theta,\beta)} \Gamma\left(1+\frac{1}{\theta}\right).
\end{split}
\eeq
By using the normalizing constant $Z$ in \eqref{eqn:Z11} and the variance in \eqref{eqn:sigma11} as well as the property that $\Gamma(1+z)=z\Gamma(z)$ for $z>0$, the entropy power can be written in terms of the variance as follows
\beq
\begin{split}
e^{2h(p)}
%=&4c^{-\frac{2}{\theta}}\theta^{-2}\left(\Gamma\left(\frac{1}{\theta}\right)\right)^2\exp\left\{\frac{2\Gamma\left(1+\frac{1}{\theta}\right)}{\Gamma\left(\frac{1}{\theta}\right)}\right\}  \\
=&4\beta^{-\frac{2}{\theta}}\theta^{-2}\left(\Gamma\left({1}/{\theta}\right)\right)^2e^{{2}/{\theta}} \\
=&A(\theta)\cdot\text{Var}(X)
\end{split}
\eeq
where
\beq
A(\theta)=4\theta^{-2}\frac{\left(\Gamma\left({1}/{\theta}\right)\right)^3}{\Gamma\left({3}/{\theta}\right)}e^{{2}/{\theta}}.
\eeq

\section{Proof of Lemma~\ref{lem:existence}}\label{sec:proof_lem_sec5}
Suppose that $p(\cdot)$ satisfies the Lipschitz condition with constant $c_s>0$, i.e.,
\beq\label{eqn:Lipschitz_cl}
|p(x+y)-p(x)| \leq c_s |y|
\eeq
for any $x, y$. 
From the  Lipschitz continuity of $p(x)$ and the fact that there exists $x\in[kh,(k+1)h)$ such that $\bar{p}_n(x)=p(x)$ for every $k\in\{-n,\dots,n-1\}$,  the difference between $p(x)$ and $\bar{p}_n(x)$  can be bounded as 
\beq\label{eqn:pbarp_bd}
|p(x)-\bar{p}_n(x)|\leq c_s  h
\eeq
for any $x\in[kh,(k+1)h)$. Let $M_k:=\max_{x\in[kh,(k+1)h)}p(x)$, then
\beq\label{eqn:bdmkcl}
|p(x)-\bar{p}_n(x)|\leq \min\{M_k, c_s h\}
\eeq
for every $x\in[kh,(k+1)h)$, since $p(x),\bar{p}_n(x)\geq 0$ for any $x\in\mathbb{R}$.

The difference in variances of $p(x)$ and of $\bar{p}_n(x)$ can be bounded as
\beq
\begin{split}\label{eqn:bd_var_n}
&\left|\int_{-s}^s x^2 p(x) dx-\int_{-s}^s x^2 \bar{p}_n(x) dx \right|\\
&\leq \int_{-s}^s x^2|p(x)-\bar{p}_n(x)|dx\\
&=2\int_{0}^s x^2|p(x)-\bar{p}_n(x)| dx\\
&=2\sum_{k=0}^{n-1} \int_{kh}^{(k+1)h} x^2|p(x)-\bar{p}_n(x)| dx\\
&\leq 2 c_s h \sum_{k=0}^{n-1} \int_{kh}^{(k+1)h} x^2 dx=\frac{2}{3}c_s h  s^3=\frac{2 c_s s^4}{3} \frac{1}{n}
\end{split}
\eeq
where $h=s/n$. The last inequality in~\eqref{eqn:bd_var_n} is from~\eqref{eqn:pbarp_bd}. 

Consider next the difference in entropies of $p(x)$ and of $\bar{p}_n(x)$, denoted by $h(p)$ and $h(\bar{p}_n)$, respectively.
Define $f(y)=-y\log y$ for $y>0$ with $f(0)=0$.
We approximate $f(p(x))$ for any $x\in [kh,(k+1)h)$ by a Taylor expansion at $\bar{p}_n(x)=\frac{1}{h}\int_{kh}^{(k+1)h} p(x) dx$,
\beq
\begin{split}\label{eqn:fp_Taylor}
f(p(x))
=&f(\bar{p}_n(x))+\frac{df(y)}{dy}\Big|_{y=\bar{p}_n(x)}(p(x)-\bar{p}_n(x))+O((p(x)-\bar{p}_n(x))^2/\bar{p}_n(x))\\
=&-\bar{p}_n(x)\log \bar{p}_n(x)-(1+\log \bar{p}_n(x))(p(x)-\bar{p}_n(x))+O((p(x)-\bar{p}_n(x))^2/\bar{p}_n(x)).
%\leq &-\bar{p}_n(x)\log \bar{p}_n(x)+\frac{c_1' s}{n}
\end{split}
\eeq
Thus,
\beq\label{eqn:fp_Taylor_er}
\begin{split}
|f(p(x))-f(\bar{p}_n(x))|\leq&|1+\log\bar{p}_n(x)|\cdot|p(x)-\bar{p}_n(x)|+O((p(x)-\bar{p}_n(x))^2/\bar{p}_n(x)).
\end{split}
\eeq
%for a positive constant $c_1'$ where the last inequality is from the fact $\log\bar{p}_n(x)$ is bounded by a constant.
%Note that $\bar{p}_n(x)>l>0$ for some constant $l$ since $p(x)$ is assumed to have a compact support in $[-s,s]$ and $\bar{p}_n(x)$ in $x\in[kh,(k+1)h)$ is equal to the average value of $p(x)$ in each sub-interval $[kh,(k+1)h)$.
To further bound  the right hand side of this inequality, we next find an upper bound on $|1+\log \bar{p}_n(x)|$ for $x\in[kh,(k+1)h)$. 
Note that $\bar{p}_n(x)$ for $x\in[kh,(k+1)h)$ is equal to the area of $p(x)$ over $[kh,(k+1)h)$ divided by $h$, i.e., $\bar{p}_n(x)=\frac{1}{h}\int_{kh}^{(k+1)h} p(x)dx$.
Since $p(x)$ is a unimodal distribution, in each sub-interval $ [kh,(k+1)h)$, $p(x)$ is either monotonically increasing (when $-n\leq k\leq -1$) or monotonically decreasing (when $0\leq k\leq n-1$). 
Let us consider the case when $-n\leq k\leq -1$. 
For this case $M_k=\max_{x\in[kh,(k+1)h)}p(x)=p((k+1)h)$. From the Lipschitz continuity of $p(x)$, the maximum slope of $p(x)$ is smaller than $c_s$. Therefore,   for $x\in[kh,(k+1)h)$ with $-n\leq k\leq -1$ we can bound $p(x)$ as
\beq
p(x)\geq \max\{c_s(x-kh)+M_k-c_sh,0\}.
\eeq 
The area under $p(x)$ in the interval $[kh,(k+1)h)$ is thus lower bounded by the area under the curve $ \max\{c_s(x-kh)+M_k-c_sh,0\}$ over $x\in[kh,(k+1)h)$, i.e.,
\beq\label{eqn:area_bd}
\begin{split}
&\bar{p}_n(x)=\frac{1}{h}\int_{kh}^{(k+1)h} p(x) dx\\
&\geq\begin{cases}
 \left(M_k-\frac{c_s h}{2}\right)\geq \frac{c_s h}{2},&\text{when} \;M_k\geq c_sh,\\
\frac{1}{h}\frac{M_k^2}{2c_s},&\text{when} \; M_k<c_sh.
\end{cases}
\end{split}
\eeq
When $M_k\geq c_s h$, from~\eqref{eqn:area_bd} and  $h=s/n$, it can be shown that $\frac{c_s s}{2n}\leq \bar{p}_n(x)=\frac{1}{h}\int_{kh}^{(k+1)h}p(x)dx\leq \frac{1}{h}=\frac{n}{s}$, which implies $|1+\log\bar{p}_n(x)|\leq c\log n$ for some constant $c>0$.
Combining this bound with $|p(x)-\bar{p}_n(x)|\leq  \frac{c_s s}{n}$ from~\eqref{eqn:bdmkcl}, we obtain
\beq\label{eqn:err_b1}
|1+\log\bar{p}_n(x)|\cdot|p(x)-\bar{p}_n(x)|\leq \frac{c'\log n}{n}
\eeq
for some constant $c'>0$.

When $M_k< c_s h$, from~\eqref{eqn:bdmkcl} it can be shown that $|p(x)-\bar{p}_n(x)|\leq M_k<\frac{c_ss}{n}$, and from~\eqref{eqn:area_bd} it can be shown that $\frac{M_k^2 n}{2c_ss}\leq \bar{p}_n(x)=\frac{1}{h}\int_{kh}^{(k+1)h}p(x)dx\leq\frac{1}{h}= \frac{n}{s}$. %, which implies $|1+\log\bar{p}_n(x)|\leq c\log n$ for some constant $c>0$.
Since $|M_k\log M_k^2|\leq \frac{2c_ss}{n}\log\frac{c_s s}{n}$ for $M_k\leq \frac {c_s s}{n}\leq e^{-1}$ with a sufficiently large $n$,
\beq\label{eqn:err_b2}
|1+\log\bar{p}_n(x)|\cdot|p(x)-\bar{p}_n(x)|\leq\frac{c''\log n}{n}
\eeq
for some constant $c''>0$.
Therefore, by using~\eqref{eqn:err_b1} and~\eqref{eqn:err_b2}, we can bound $|f(p(x))-f(\bar{p}_n(x))|$ in~\eqref{eqn:fp_Taylor_er} as
\beq
|f(p(x))-f(\bar{p}_n(x))|\leq \frac{c\log n}{n}+O(n^{-1})
\eeq
for some constant $c>0$. Even though we proved this bound for $x\leq 0$, i.e, for $x\in[kh,(k+1)h)$ with $-n\leq k\leq -1$, due to symmetry of $p(x)$ about $x=0$, this bound also holds for any $x\geq 0$.
Therefore, the difference in entropies of $p(x)$ and $\bar{p}_n(x)$ can be bounded as
\beq
\begin{split}
&|h(p)-h(\bar{p}_n)|\\
&=\Big|\int_{-s}^{s}\left(-p(x)\log p(x)+\bar{p}_n(x)\log\bar{p}_n(x)\right)dx  \Big|\\
&=\Big|\sum_{k=-n}^{n-1}\int_{kh}^{(k+1)h} (f(p(x))-f(\bar{p}_n(x)))dx \Big|\\
&\leq \sum_{k=-n}^{n-1}\int_{kh}^{(k+1)h}|f(p(x)-f(\bar{p}_n(x))|dx\\
&\leq \sum_{k=-n}^{n-1} h\frac{c\log n}{n}+O(n^{-1})\\
&= 2ch\log n+O(n^{-1})=\frac{2cs\log n}{n}+O(n^{-1}).
\end{split}
\eeq
%to bound the difference between $f(p(x))$ and $-\bar{p}_n(x)\log \bar{p}_n(x)$.
%Note that $\bar{p}_n(x)$ is the are mean value of $p(x)$ in each sub-interval $[kh, (k+1)h)$.
This bound then implies that
\beq
\begin{split}
e^{2h(\bar{p}_n)}\leq e^{2h(p)}(1+c_2n^{-1}\log n)
\end{split}
\eeq
for some constant $c_2>0$.

\section{Proof of Lemma~\ref{lem:bd_linear_mix_new}}\label{sec:proof_var_bd_linear_mix_new}
Suppose that $p(x)$ satisfies the Lipschitz condition with constant $c_s>0$, i.e.,
\beq
|p(x+y)-p(x)| \leq c_s |y|
\eeq
for any $x, y$. From~\eqref{eqn:alphai_or}, for $t=1,\dots, n$,
\beq \begin{split}
\alpha_t &=  2t \int_{-(t+1)h}^{-th} \left( p(x+h) - p(x) \right) dx \\
&\leq 2 c_s t h^2,
\end{split} \eeq
where $h = s/n$. We also have the relation $\sum_{t=1}^n \alpha_t = 1$.

For the linear mixture density $\bar p_n(x) $ in~\eqref{eqn:pnxalphai}, 
$$
\bar p_n(x) = \sum_{t=1}^n \alpha_t\cdot \mathrm{unif} (-th, th),
$$
the variance of $X\sim \bar{p}_n(x)$ is
\beq
\begin{split}\label{var11}
&\var(\bar{p}_n) = \frac{1}{3} \sum_{t=1}^n \alpha_t (th)^2 = \frac{s^2}{3} \sum_{t=1}^n \alpha_t \left( \frac{t}{n} \right)^2 \\
&\leq\frac{2 c_s s^4}{3} \sum_{t=1}^n \frac{t^3}{n^4} = \frac{c_s s^4}{6} + O(n^{-1}),
\end{split}
\eeq
and the entropy of $X\sim \bar{p}_n(x)$ is
\beq \begin{split}\label{eqn:87}
h(\bar p_n) &\geq \sum_{t=1}^n \alpha_t \log (2th) = \log(2s) + \sum_{t=1}^n \alpha_t \log \frac{t}{n} \\
&\geq \log(2s) + \sum_{t=1}^n 2 c_s t h^2 \log \frac{t}{n} \\
&=\log(2s) + 2 c_s s^2 \sum_{t=1}^n \frac{t}{n^2} \log \frac{t}{n} \\
&= \log(2s) - \frac{c_s s^2}{2} + O(n^{-1}),
\end{split} \eeq
where the last equality holds since
$$
\sum_{t=1}^n \frac{t}{n^2} \log \frac{t}{n} = \int_0^1 x \log x \, dx + O(n^{-1}) = -\frac{1}{4} + O(n^{-1}).
$$
From~\eqref{eqn:87}, the entropy power $e^{2h(\bar p_n)}$ is bounded below as
\beq\label{en11}
e^{2h(\bar p_n)} \geq \frac{4s^2}{e^{c_s s^2}} \left( 1 - c_3' n^{-1} \right)
\eeq
for some constant $c_3' > 0$. 

From~\eqref{var11} and~\eqref{en11}, we obtain
\beq
\var(\bar p_n) \leq \frac{c_s s^2 e^{c_s s^2}}{24} e^{2h(\bar p_n)} \left( 1 + c_3 n^{-1} \right)
\eeq
for some constant $c_3 > 0$.

\section{Proof of Lemma~\ref{lem:bound_r_v}}\label{app:lem:bound_r_v}
In Lemma~\ref{lem:bound_r_v}, we show that the parameter $r_v=\max\{\var(p_l)/\var(p_r),\var(p_r)/\var(p_l)\}$ for Lipschitz-continuous symmetric unimodal densities $p_l$ and $p_r$ in~\eqref{eqn:plprdef}, defined as a function of a Lipschitz-continuous unimodal density $p(x)$ over bounded support $[b-s_l,b+s_r]$ with the unique mode at $x=b$, is bounded above as
%Define $r_v=\max\{\var(p_l)/\var(p_r),\var(p_r)/\var(p_l)\}$ for Lipschitz-continuous symmetric unimodal densities $p_l$ and $p_r$ defined in~\eqref{eqn:plprdef} as a function of Lipschitz-continuous unimodal density $p$. The parameter $r_v$ is bounded as
\beq
r_v\leq 128(c_s s^2)^4
\eeq
where $s=\max\{s_l,s_r\}$ and $c_s$ is the Lipschitz constant of $p(x)$ such that
\beq
|p(x)-p(x+y)|\leq c_s|y|.
\eeq
From the definition of the two densities $p_l$ and $p_r$ in~\eqref{eqn:plprdef}, the variances $\var(p_l)$ and $\var(p_r)$ of the respective densities are
\beq
\begin{split}
\var(p_l)=&\frac{1}{\beta_l} \int_{b-s_l}^b (x-b)^2 p(x) dx,\\
\var(p_r)=&\frac{1}{\beta_r} \int_{b}^{b+s_r} (x-b)^2 p(x) dx,
\end{split}
\eeq
where $\{\beta_l,\beta_r\}$ denote the areas under the density $p(x)$  in the left and right of the mode $x=b$, respectively, as defined in~\eqref{eqn:blbrarea}

To find an upper bound on  $r_v=\max\{\var(p_l)/\var(p_r),\var(p_r)/\var(p_l)\}$, we first find an upper bound on $r_\beta:=\max\{\beta_l/\beta_r,\beta_r/\beta_l\}$.
Assume without loss of generality that $\beta_l \leq 1/2 \leq \beta_r$. Let $t = p(b)$. Then, for any $x < b$, we have from the Lipschitz condition $|p (b) - p (x)| \leq c_s |b-x|$ that
\beq
	p (x) \geq t - c_s |b-x|.
\eeq
Integrating the above inequality, we find that
\beq \label{eq:beta_1_lower}
	\beta_l = \int_{-\infty}^b p (x) dx \geq \int_{b-t/c_s}^b p (x) dx = \frac{t^2}{2c_s}.
\eeq
It can be also shown that $s_r t \geq \beta_r$ from the unimodality condition. Hence,
\beq\label{eqn:up_bd_r_b}
	r_\beta=\frac{\beta_r}{\beta_l} \leq \frac{2\beta_r c_s}{t^2} \leq \frac{2c_s s_r^2}{\beta_r} \leq 4 c_s s_r^2\leq 4c_s s^2.
\eeq

We next find an upper bound on
\beq\label{eqn:r_i}
r_i:=\max\left\{\frac{\int_{b-s_l}^b (x-b)^2 p(x) dx}{ \int_{b}^{b+s_r} (x-b)^2 p(x) dx}, \frac{ \int_{b}^{b+s_r} (x-b)^2 p(x) dx}{\int_{b-s_l}^b (x-b)^2 p(x) dx}\right\}.
\eeq
Since $p(x)$ has the unique mode at $x=b$, when we let $t=p(b)$, $p(x)$ satisfies
\beq
t-c_s|b-x|\leq p(x)\leq t
\eeq
for any $x\in[b-s_l,b+s_r]$.
By using this bound, we can show that
\beq
\begin{split}
\frac{t^4}{12c_s^3} \leq& \int_{b-s_l}^b (x-b)^2 p(x) dx\leq \frac{t s_l^3}{3},\\
\frac{t^4}{12c_s^3} \leq& \int_{b}^{b+s_r} (x-b)^2 p(x) dx\leq \frac{t s_r^3}{3}.
\end{split}
\eeq
Therefore, the parameter $r_i$ in~\eqref{eqn:r_i} is bounded above as
\beq
r_i\leq \frac{4(c_s s)^3}{t^3}
\eeq
for $s=\max\{s_l,s_r\}$. Since $p(x)$ has the unique mode at $x=b$ with the value $t=p(b)$, $t$ is bounded below as $t\geq 1/(2s)$. Hence,
\beq\label{eqn:up_bd_r_i}
r_i\leq 32 (c_s s^2)^3.
\eeq
By using~\eqref{eqn:up_bd_r_b} and~\eqref{eqn:up_bd_r_i} and from the definition of $r_v$, it can be shown that
\beq
\begin{split}
r_v\leq r_\beta r_i\leq 128 (c_ss^2)^4.
\end{split}
\eeq  
\bibliographystyle{IEEEtran}
%\bibliography{IEEEabrv,Superadd_v3}

\end{document}